\documentstyle[12pt,epsfig]{article}
\newcommand{\be}{\begin{equation}}
\newcommand{\ee}{\end{equation}}
\newcommand{\bea}{\begin{eqnarray}}
\newcommand{\eea}{\end{eqnarray}}

\newcommand{\ggam}{\gamma\gamma}

\newcommand{\beq}{\begin{equation}}
\newcommand{\eeq}{\end{equation}}

\newcommand{\nn}{\nonumber}

\def\rhosq{\sqrt{\frac{s-4m^2}s}}

\def\ga{\mathrel{\mathpalette\fun >}}
\def\fun#1#2{\lower3.6pt\vbox{\baselineskip0pt\lineskip.9pt
\ialign{$\mathsurround=0pt#1\hfil##\hfil$\crcr#2\crcr\sim\crcr}}}

\begin{document}

\title{Quark--antiquark
states and their radiative transitions in terms of the spectral
integral equation.\\
{\Huge I.} Bottomonia}
\author{V.V. Anisovich, L.G. Dakhno, M.A. Matveev,\\ V.A. Nikonov
and A.V. Sarantsev}

\date{\today}
\maketitle

\begin{abstract}
In the framework of the spectral integral equation,
we consider the $b\bar b$ states  and  their radiative transitions.
We reconstruct the $b\bar b$ interaction on the basis of
data for the levels of the bottomonium states with $J^{PC}=0^{-+}$,
$1^{--}$, $0^{++}$, $1^{++}$, $2^{++}$
as well as the data for the radiative transitions
$\Upsilon(3S) \to\gamma\chi_{bJ}(2P) $ and
$\Upsilon(2S) \to\gamma \chi_{bJ}(1P) $ with $J=0,1,2$.
We calculate  bottomonium  levels with the radial quantum numbers
$n\le 6$, their wave functions and corresponding radiative transitions.
The ratios
$Br[\chi_{bJ}(2P)\to\gamma\Upsilon(2S)]/Br[\chi_{bJ}(2P)\to\gamma\Upsilon(1S)]$
for $J=0,1,2$ are found in the agreement with data.
We determine the $b\bar b$ component of the photon wave function
using the data for the $e^+e^-$ annihilation,
$e^+e^- \to\Upsilon( 9460)$, $\Upsilon(10023)$, $\Upsilon(10036)$,
$\Upsilon(10580)$, $ \Upsilon(10865)$, $\Upsilon(11019)$,
and predict partial widths of the two-photon decays
$\eta_{b0}\to\ggam$, $\chi_{b0}\to\ggam$, $\chi_{b2}\to\ggam$
for the radial excitation states below $B\bar B$ threshold ($n\le 3$).
\end{abstract}

\section{Introduction}

In \cite{BS}, see also \cite{book}, the program was suggested for the
reconstruction of soft quark--antiquark interaction, on the basis of
data on  meson levels and meson radiative transitions. Now, we have
the first results in the realization of this program for the $b\bar
b$, $c\bar c$ systems and light quarkonia $q\bar q$. In this paper we
present the results for the bottomonia.

In \cite{BS}, the
equations for  $q\bar q$ systems were written in terms of the spectral
integral representation. The spectral integration
technique is precisely advantageous for  composite particles, for the
content of a composite system is thus strictly controlled and there is
no problem with the description of  high spin states. The equation
for the composite $q\bar q$ system in the spectral integration
technique \cite{BS} is a direct generalization of the dispersion
$N/D$ equation, when we represent the $N$-function  as a sum of
separable vertices. In \cite{BS}, this equation was conventionally
called  the spectral integral Bethe--Salpeter equation. However, it
should be emphasized that in certain important points it differs from
the standard Bethe--Salpeter equation \cite{bethe} written in the
Feynman technique (the application of the Feynman technique to the
calculation of meson states may be found, for example, in
\cite{faustov,petry-meson,Linde,Resag,Kuhn}
and references therein).

The strong QCD is responsible for the formation of composite systems,
though, despite a remarkable progress in this field, certain
features of the soft interaction of color objects are
rather enigmatic till now. A fascinating feature of the light quark
($q=$ $u$, $d$, $s$) meson spectra
is the linearity of their trajectories on both  the $(J,M^2)$- and
$(n,M^2)$-planes ($J$ is the spin of the $q\bar q$ system with
the mass $M$ and  $n$ is its radial quantum number)
 \cite{book,syst}.
The linearity of trajectories has been also seen
in  the baryon sector \cite{book,klempt} --- this fact allows one to
suggest a specific role of diquarks in baryon systems, see the
discussion in \cite{book,klempt} and references therein.
Unfortunately, the number of experimentally observed excited states for
baryons is much less than that given by the standard quark model
calculation (e.g., see \cite{petry-baryon}) that limits our
understanding of the  three-quark interactions.

In practice, all the observed highly excited meson states (with
$M>1500$ MeV) are lying on linear $q\bar q$ trajectories, and we have
no candidates for the $q\bar q g$ hybrids. The effective gluon $g$
has the mass $m_g$ of the order of $m_g \simeq 700-1000$ MeV,
\cite{gluon,cornwell,gerasyuta,g-lat},
so the hybrid mesons could be spread
over the region $1500-2000$ MeV. Still, the experiment does
not point to the increase of the meson state density in this region.
Also, there are no definite indications to the existence of mesons with
the exotic quantum numbers inherent in hybrids \cite{amsler}.

The particulars of hadron mass spectra discussed above point to the fact
that in studying mesons in the soft region (or formed at large
distances) the most reliable way is to
consider the quark--antiquark interaction  by determining their
characteristics from the data. It is the guideline in
our approach to the quark--antiquark systems.

The spectral integral equation \cite{BS} gives us a unique
solution for the quark--antiquark levels and their wave functions,
provided the interquark interaction is known. Let us emphasize
that the equations work for both instantaneous interactions (or those
of the potential type) and the  $t$-channel exchanges with retardation,
 and even
for the energy-dependent interactions: this  follows from the fact
that the equations themselves are the modified dispersion
relations for the amplitude. For solving the inverse problem, that
is, for  reconstructing the interaction, it is not enough to know the
meson masses --- one should know the wave functions of
quark--antiquark
systems. Such an information is contained  in the hadronic form
factors and partial widths of radiative decays. Therefore, in the
present approach, we consider simultaneously the meson spectra in terms
of the spectral integral equations and meson radiative transitions in
terms of the dispersion relations over meson masses --- in this way,
all the calculations are carried out within compatible methods.

The method of calculation of the radiative transition amplitudes in
terms of the double dispersive integrals was developed in a number of
papers
\cite{deut,physrev,epja}.
An important point was the
representation of the transition amplitude in the form convenient for
simultaneous fitting to the spectral integral equation
 --- it was done in \cite{YFscalar,YFtensor}.

A significant information on the
quark--antiquark meson wave functions is hidden in the two-photon
meson decays: $meson\to \gamma\gamma$. For the calculation of such
processes, one needs to know the quark wave function of
the photon; the method of reconstruction of the
$\gamma\to q\bar q$ and $\gamma\to c\bar c$ vertices was developed in
\cite{PR-g,YF-g}.

The paper is organized as follows.

In Section 2, we present the
technique and basic formulas which were used in  fitting to
quark--antiquark states. Here, we briefly recall the  spectral integral
equation and give the formulas for the
radiative transition amplitudes
 of quark systems  calculated
within the double spectral integrals. Keeping in mind the application
of these formulas to the other quark systems, we do not specify the
flavor of the considered quark:
$(Q\bar Q)_{in}\to \gamma+(Q\bar Q)_{out}$ ,
$e^+e^- \to V(Q\bar Q)$  and $Q\bar Q$-$meson\to \gamma  \gamma $.

In Section 3, we consider the
bottomonium systems: the choice of the $b\bar b$-mesons as
a primary object for the study is motivated by the
large $b$-quark mass, so the nonrelativistic approximation is expected
to work well, and we can reliably compare our results with those
obtained within  nonrelativistic approaches
\cite{Hulth,Godfrey,Gupta,Lucha}.
Different variants of the interaction
(instantaneous and retarded) are discussed. The results of fitting to
masses and radiative transitions  (for both observed and
predicted states) are presented and discussed. In the $b\bar b$ sector,
we reconstruct the interaction, on the basis of data on the levels of
bottomonia with $J^{PC}=0^{-+}$, $1^{--}$, $0^{++}$, $1^{++}$, $2^{++}$
and radiative transitions
$\Upsilon(3S) \to\gamma \chi_{bJ}(2P) $ and
$\Upsilon(2S) \to\gamma \chi_{bJ}(1P) $ at $J=0,1,2$.
We calculate the bottomonium  levels
with radial quantum numbers
$n\le 6$, as well as their wave functions and corresponding radiative
transitions.
 We determine the $b\bar b$ component of the photon wave function on
 the basis of data for the $e^+e^-$ annihilation reactions $e^+e^- \to
\Upsilon( 9460)$,$\Upsilon(10023)$,$\Upsilon(10036)$,
$\Upsilon(10580)$,$ \Upsilon(10865)$,$\Upsilon(11019)$,
and predict partial widths of the two-photon decays
$\eta_{b0}\to\ggam$, $\chi_{b0}\to\ggam$,
$\chi_{b2}\to\ggam$ for the excited states with $n\le 3$.

 Brief summary  is given in Conclusion.

\section{Spectral integral equation and radiative
transition amplitudes}
Here we present the formulas used in fitting to the
$b\bar b$ systems. Of course, they may be also applied for the
charmonia, as well as for light quark states $q\bar q$ with $I=1$, or
one-flavor states with $I=0$ (pure $s\bar s$ or
$n\bar n=(u\bar u+d\bar d)/\sqrt{2}$ systems.

\subsection{Spectral integral  equation}

The  wave function of the quark--antiquark meson with the mass $M$
is characterized by the total momentum $J$,
quark--antiquark spin $S$ (in the flavor nonet with fixed $J^P$, the
values $S=0,1$ determine $C$ parity) and radial  number $n$. We
denote the wave function
as $\widehat\Psi^{(S,J)}_{(n)\,\mu_{1}\cdots\mu_{J}} (k_{\perp})$,
with $k_{\perp}$ being relative quark momentum and the indices
$\mu_{1}, _{\cdots} ,\mu_{J}$  related to the total
momentum. For the heavy-quark $Q \bar Q$ system,
the spectral integral equation reads \cite{BS}:
\be
\left(s-M^2\right)
\widehat\Psi^{(S,J)}_{(n)\,\mu_{1}\cdots\mu_{J}} (k_{\perp})=
\label{bs1}
\ee
$$
=\int \limits_{4m^2}^{\infty}\frac{ds'}{\pi}
\int d\Phi_2(P';k'_1 ,k'_2)\,
\widehat V\left(s,s',(k_{\perp}k'_{\perp})\right)(\hat k_1'+m)
\widehat \Psi^{(S,J)}_{(n)\,\mu_{1}
\cdots\mu_{J}} (k'_{\perp})(-\hat k_2'+m)\ .
$$
Here, the
quarks are mass-on-shell, $k_1^2=k_1'^2=k_2^2=k_2'^2=m^2$.
The phase space factor in the intermediate state is determined
as follows:
\be
d\Phi_2(P';k'_1 ,k'_2) = \frac 12 \frac{d^3k'_1}{(2\pi)^3\, 2k'_{10}}
\frac{d^3k'_2}{(2\pi)^3\, 2k'_{20}}
(2\pi)^{4}\delta^{(4)}(P'-k'_1 -k'_2)\ .
\ee
The following notations are used:
\be
k_\perp=\frac12\left(k_1-k_2\right)\ , \quad
P=k_1+k_2\ ,
\qquad k'_\perp=\frac12\left(k'_1-k'_2\right)\ , \quad
P'=k'_1+k'_2\ ,
\label{bs2}
\ee
$$
P^2=s\  ,\quad \ P'^2=s'\  , \qquad
g^{\perp}_{\mu\nu} =g_{\mu\nu}-
\frac{P_\mu P_\nu}{s} \  , \quad
g'^{\perp}_{\mu\nu} =g_{\mu\nu}-
\frac{P'_\mu P'_\nu}{s'} \  ,
$$
so one can write
$\ k^{\perp}_{\mu}=k_{\nu}g^{\perp}_{\nu\mu}$ and
$\ k'^{\perp}_{\mu}=k'_{\nu}g'^{\perp}_{\nu\mu}$\ .
In the center-of-mass system, the integration  can be
re-written as
\be
\int \limits_{4m^2}^{\infty}\frac{ds'}{\pi}\int d\Phi_2(P';k'_1
,k'_2)\longrightarrow \int \frac{d^3k'}{(2\pi)^3k'_0}\ ,
\label{int_tr}
\ee
where $k'$ is the momentum of one of the quarks.

For the fermion--antifermion
system with definite $J,S$ and $L$ (angular momentum), we introduce
the moment operators
 $Q^{(S,L,J)}_{\mu_{1}\cdots\mu_{J}}(k_{\perp})$
 defined as follows \cite{operators}:
\bea
\label{bs10}
&& Q^{(0,J,J)}_{\mu_1\mu_2\ldots\mu_J}(k_{\perp})=
i\gamma_5
X_{\mu_1\ldots\mu_J}(k^\perp)
\sqrt{\frac{2J+1}{\alpha_J}}  \ ,
\\
&& Q^{(1,J,J)}_{\mu_1\ldots\mu_J}(k_{\perp})=
\frac{i\varepsilon_{\alpha\eta\xi\gamma}\gamma_\eta
k^\perp_\xi P_\gamma Z^\alpha_{\mu_1\ldots\mu_J}}
{\sqrt s}
\sqrt{\frac{(2J+1)J}{(J+1)\alpha_J}}\ ,
\nonumber
\\
&& Q^{(1,J+1,J)}_{\mu_1\ldots\mu_J}(k_{\perp})=
\gamma_\alpha X_{\alpha\mu_1\ldots\mu_J}
\sqrt{\frac{J+1}{\alpha_J}}\ ,
\nonumber
\\
&& Q^{(1,J-1,J)}_{\mu_1\ldots\mu_J}(k_{\perp})=
\gamma_\alpha Z^{\alpha}_{\mu_1\ldots\mu_J}
\sqrt{\frac{J}{\alpha_J}} \ ,
\eea
where $\alpha_J=(2J-1)!!/J!$ and
\bea
&&X^{(J)}_{\mu_1\ldots\mu_J}(k_\perp)
=\frac{(2J-1)!!}{J!}
\bigg [
k^\perp_{\mu_1}k^\perp_{\mu_2}k^\perp_{\mu_3}k^\perp_{\mu_4}
\ldots k^\perp_{\mu_J} -
\label{x-direct}
\\
&&-\frac{k^2_\perp}{2J-1}\left(
g^\perp_{\mu_1\mu_2}k^\perp_{\mu_3}k^\perp_{\mu_4}\ldots k^\perp_{\mu_J}
+g^\perp_{\mu_1\mu_3}k^\perp_{\mu_2}k^\perp_{\mu_4}\ldots
k^\perp_{\mu_J} + \ldots \right)+
\nonumber
\\
&&
+\frac{k^4_\perp}{(2J-1)
(2J-3)}\left(
g^\perp_{\mu_1\mu_2}g^\perp_{\mu_3\mu_4}k^\perp_{\mu_5}
k^\perp_{\mu_6}\ldots k^\perp_{\mu_J}+
\right .
\nonumber
\\
&&
\left .
+g^\perp_{\mu_1\mu_2}g^\perp_{\mu_3\mu_5}k^\perp_{\mu_4}
k^\perp_{\mu_6}\ldots k^\perp_{\mu_J}+
\ldots\right)+\ldots\bigg ]\ ,
\nonumber
\eea
\bea
Z^{(J-1)}_{\mu_1\ldots\mu_J, \alpha}(k_\perp)&=&
\frac{2J-1}{J^2}\left (
\sum^J_{i=1}X^{{(J-1)}}_{\mu_1\ldots\mu_{i-1}\mu_{i+1}\ldots\mu_J}
(k_\perp)g^\perp_{\mu_i\alpha} -\right .
\nonumber \\
&&\left . -\frac{2}{2J-1} \sum^J_{i,j=1 \atop i<j}
g^\perp_{\mu_i\mu_j}
X^{{(J-1)}}_{\mu_1\ldots\mu_{i-1}\mu_{i+1}\ldots\mu_{j-1}\mu_{j+1}
\ldots\mu_J\alpha}(k_\perp) \right )\ .
\nonumber
\eea
The operators are normalized as:
\bea
&&\hspace{-0.5cm}
\int\frac{d\Omega}{4\pi}{\rm Sp}[Q^{(0,J,J)}_{\mu_1\ldots\mu_L}(m+\hat k_1)
Q^{(0,J,J)}_{\nu_1\ldots\nu_L}(m-\hat k_2)]=-2s{\bf k}^{2J}
(-1)^J O^{\mu_1\ldots\mu_J}_{\nu_1\ldots\nu_J}\  ,
\nonumber \\
&&\hspace{-0.5cm}
\int\frac{d\Omega}{4\pi}{\rm Sp}[Q^{(1,J,J)}_{\mu_1\ldots\mu_J}(m+\hat k_1)
Q^{(1,J,J)}_{\nu_1\ldots\nu_J}(m-\hat k_2)]=-2s{\bf k}^{2J}
(-1)^J O^{\mu_1\ldots\mu_J}_{\nu_1\ldots\nu_J}\ ,
\nonumber \\
&&\hspace{-0.5cm}
\int\frac{d\Omega}{4\pi}{\rm Sp}[Q^{(1,J+1,J)}_{\mu_1\ldots\mu_n}(m+\hat k_1)
Q^{1,J+1,J}_{\nu_1\ldots\nu_J}(m-\hat k_2)]=
\Big (\frac{8(J+1){\bf k}^2}{2J+1}-2s\Big){\bf k}^{2(J+1)}
(-1)^J O^{\mu_1\ldots\mu_J}_{\nu_1\ldots\nu_J}\ ,
\nonumber \\
&&\hspace{-0.5cm}
\int\frac{d\Omega}{4\pi}{\rm Sp}[Q^{(1,J-1,J)}_{\mu_1\ldots\mu_J}(m+\hat k_1)
Q^{(1,J-1,J)}_{\nu_1\ldots\nu_J}(m-\hat k_2)]=
\Big ( \frac{8J{\bf k}^2}{2J+1}-2s\Big){\bf k}^{2(J-1)}
(-1)^J O^{\mu_1\ldots\mu_J}_{\nu_1\ldots\nu_J}\ ,
\nonumber \\
&&\hspace{-0.5cm}
\int\frac{d\Omega}{4\pi}{\rm Sp}[Q^{(1,J-1,J)}_{\mu_1\ldots\mu_J}(m+\hat k_1)
Q^{(1,J+1,J)}_{\nu_1\ldots\nu_J}(m-\hat k_2)]=
-8\frac{\sqrt{J(J+1)}}{2J+1}{\bf k}^{2(J+1)}
(-1)^J O^{\mu_1\ldots\mu_J}_{\nu_1\ldots\nu_J}\ .
\label{v3P2_1}
\eea
Here, $O^{\mu_1\ldots\mu_n}_{\nu_1\ldots\nu_n}$ is the projection
operator to a state with the momentum $J$ and $s=4m^2+4\vec{ k}^2$.
The projection operators have the following properties:
\be
X^{(L)}_{\mu_1\ldots\mu_L}(k_\perp)
O^{\mu_1\ldots\mu_L}_{\nu_1\ldots \nu_L}\
=\ X^{(L)}_{\nu_1\ldots \nu_L}(k_\perp)\ ,
\qquad\qquad
O^{\mu_1\ldots\mu_L}_{\alpha_1\ldots\alpha_L} \
O^{\alpha_1\ldots\alpha_L}_{\nu_1\ldots \nu_L}\
=\ O^{\mu_1\ldots\mu_L}_{\nu_1\ldots \nu_L}\ .
\label{proj_op}
\ee
Let us underline that $(-1)^J O^{\mu_1\ldots\mu_J}_{\nu_1\ldots\nu_J}$
describes the spin structure of the propagator of a particle with  the
total spin $J$ (see \cite{operators} for more detail).

In terms of these operators,
the wave functions read:
\be
\widehat
\Psi^{(S,J)}_{(n)\,\mu_{1}\cdots\mu_{J}}(k_{\perp})=
Q^{(S,J,J)}_{\mu_{1}\cdots\mu_{J}}(k_{\perp})\,
\psi^{(S,L=J,J)}_n (k_{\perp}^2)\ , \qquad S=0,1\, {\rm and}\, J=L,
\label{bs3}
\ee
$$
\widehat
\Psi^{(S,J)}_{(n)\,\mu_{1}\cdots\mu_{J}}(k_{\perp})=
Q^{(S,J+1,J)}_{\mu_{1}\cdots\mu_{J}}(k_{\perp})\,
\psi^{(S,L=J+1,J)}_n (k_{\perp}^2)\
+\
Q^{(S,J-1,J)}_{\mu_{1}\cdots\mu_{J}}(k_{\perp})\,
\psi^{(S,L=J-1,J)}_n (k_{\perp}^2)\ , \qquad J\neq L ,
$$
where the functions $\psi^{(S,L,J)}_n (k_{\perp}^2)$ depend on
$k_{\perp}^2$ only (recall that in the center-of-mass system
$k_{\perp}^2 =-{\bf k}^2$).

The wave functions with $L=J$ are
normalized as follows:
\be
1=\int\frac{d^3k}{(2\pi^3)k_0}\,2s\,|{\bf k}|^{2J}
|\psi^{(S,L=J,J)}_n(k_{\perp}^2)|^2 \, ,
\label{norm_1}
\ee
while for $L=J\pm 1$ the normalization reads:
\be \label{norm_3}
1=\int\!\frac{d^3k}{(2\pi^3)k_0}
|\psi^{(S,J+1,J)}_n(k_\perp^2)|^2
\Big ( 2s\!-\!\frac{8(J+1){\bf k}^2}{2J+1}\Big){\bf k}^{2(J+1)}+
\ee
$$
+16\frac{\sqrt{J(J+1)}}{2J+1}{\bf k}^{2(J+1)}
\psi^{(S,J+1,J)}_n (k_\perp^2)\psi^{*(S,J-1,J)}_n(k^2)
+ |\psi^{(S,J-1,J)}_n(k_\perp^2)|^2 \Big (
2s\!-\!\frac{8J{\bf k}^2}{2J+1}\Big){\bf k}^{2(J-1)}
 .  \nonumber
$$
We re-write this normalization condition as follows:
\be  \label{W}
 W_{J+1,J+1}+W_{J+1,J-1}
+W_{J-1,J-1}=1 ,
\ee
where  $W_{J+1,J+1}$, $W_{J+1,J-1}$ and
$W_{J-1,J-1}$ are determined by the wave function convolutions
$(\psi^{(S,J+1,J)}_n \psi^{(S,J+1,J)}_n)$, $(\psi^{(S,J+1,J)}_n
\psi^{(S,J-1,J)}_n)$  and $(\psi^{(S,J-1,J)}_n \psi^{(S,J-1,J)}_n)$,
correspondingly.

The interaction block can be expanded in a series with respect
to a full
set of the $t$-channel operators $\widehat O_I$:
$$ \widehat O_I ={\rm
I}, \;\; \gamma_{\mu},\;\; i\sigma_{\mu\nu},\;\; i\gamma_{\mu}\gamma_5,
\;\; \gamma_5\ \, ,
$$
\be \widehat V\left(s, s',
(k_{\perp}k'_{\perp})\right)= \sum_{I} V_I\left(s, s',
(k_{\perp}k'_{\perp})\right) \widehat O_I \otimes \widehat O_I\ ,
\label{bs4}
\ee
By solving the equation, the $t$-channel operators
should be transformed in the two $s$-channel ones; this procedure is
clarified in Appendix I.

The equation  (\ref{bs1}) is written in the
momentum representation, and we solve it in the
momentum representation  too. The equation  (\ref{bs1}) allows one
to use as an interaction the instantaneous approximation, or  take into
account the retardation effects.  In the instantaneous approximation
one has
 \be \label{2-2-9}
\widehat V\left(s,s',(k_{\perp}k'_{\perp})\right)
\longrightarrow
\widehat V(t_\perp),  \qquad
t_\perp = (k_{1\perp}-k'_{1\perp})_\mu(-k_{2\perp}+k'_{2\perp})_\mu \ .
\ee
The retardation effects are taken into account, when  the
momentum transfer squared $t$ in the interaction block depends on
the time components of the quark momentum (for more detail
see Section 2.5 of \cite{BS} and the  discussion in:
\cite{Hersbach,her2,gross,maung}):
\be   \label{2-2-10}
\widehat
V\left(s,s',(k_{\perp}k'_{\perp})\right) \longrightarrow \widehat V(t),
\qquad t = (k_{1}-k'_{1})_\mu(-k_{2}+k'_{2})_\mu\ .
\ee
Fitting to quark-antiquark states, we  use the
 interaction blocks with the following $t$-dependence:
\bea
&I_{-1}&=\frac{4\pi}{\mu^2-t} \, ,
\nonumber \\
&I_{0}&=\frac{8\pi\mu}{(\mu^2-t)^2} \, ,
\nonumber \\
&I_{1}&=8\pi\left (
\frac{4\mu^2}{(\mu^2-t)^3}-\frac{1}{(\mu^2-t)^2}
\right ),
\nonumber \\
&I_{2}&=96\pi\mu\left (
\frac{2\mu^2}{(\mu^2-t)^4}-\frac{1}{(\mu^2-t)^3}
\right ),
\nonumber \\
&I_{3}&=96\pi\left (
\frac{16\mu^4}{(\mu^2-t)^5}-\frac{12\mu^2}{(\mu^2-t)^4}+
\frac{1}{(\mu^2-t)^3}
\right )   \,  ,
\label{i0i1}
\eea
or in general case
\be
I_N=
\frac{4\pi(N+1)!}{(\mu^2-t)^{N+2}}\sum\limits_{n=0}^{N+1}
(\mu+\sqrt t)^{N+1-n}(\mu-\sqrt t)^{n}\ .
\ee
Traditionally, the interaction of heavy quarks in the instantaneous
approximation is represented in terms of the potential $V(r)$. The
 form of the  potential can be obtained  with the Fourier
transform of (\ref{i0i1}) in the center-of-mass system.
Thus, we have
\bea
t_\perp&=&-(\vec{k}-\vec{k'})^2=-\vec{q}^{\, 2}\ ,
\nonumber\\
I^{(coor)}_N(r,\mu)&=&\int{
\frac{d^3q}{(2\pi)^3}\;e^{-i\vec{q}\, \vec{r}}\;I_N(t_\perp)}\ ,
\label{icin}
\eea
that gives
\bea \label{2-2-13}
I^{(coor)}_N(r,\mu)=r^N\;e^{-\mu r}\ .
\eea
Working with the instantaneous interaction, we consider the following
types of $V(r)$:
\bea
\label{2-2-14}
V(r)= a + b\ r + c\ e^{-\mu_c\ r}+d\frac{e^{-\mu_d\ r}}{r}\ ,
\eea
where the constant and linear (confinement) terms read:
\bea
\label{2-2-15}
a&\to&a\;I^{(coor)}_0(r,\mu_{constant}\to 0)\ ,
\nonumber\\
br&\to&b\;I^{(coor)}_1(r,\mu_{linear}\to 0)\ .
\eea
The limits $\mu_{constant}\; \mu_{linear}\to 0$ mean
that in the fitting procedure
the parameters $\mu_{constant}$ and $\mu_{linear}$ are chosen to be
small enough, of the order of 1--10 MeV.  It was checked  that the
solution for the states with $n\leq 6$
is stable, when  $\mu_{constant}$ and $\mu_{linear}$ change
in this interval.

\subsubsection{Wave functions}

Let us present in the explicit form the wave functions of the
studying states:
$0^{-+}$, $1^{--}$,  $0^{++}$, $1^{++}$,  $2^{++}$, $1^{+-}$.
The wave
functions of these states,
$\hat\Psi^{(S,J)}_{(n)\mu_1,\mu_2,\cdots,\mu_J}$, in terms of the
operators  (\ref{bs10}) and invariant functions
$\psi^{(S,L,J)}_n (k_{\perp}^2)$  read:
\bea
&0^{-+}:&\quad   \widehat \Psi^{(0,0)}_{(n)} = i\gamma_5
\psi^{(0,0,0)}_n (k^2)\ , \nn\\
&1^{--}:&\quad   \widehat \Psi^{(1,1)}_{(n)\,\mu} = \gamma_\mu ^\perp
\psi^{(1,0,1)}_n (k^2)
+\frac{3}{\sqrt 2}
\left [k_\mu\hat k-\frac13 k^2\gamma^{\perp}_\mu\right ]
\psi^{(1,2,1)}_n (k^2)\ , \nn \\
&0^{++}:&\quad   \widehat \Psi^{(1,0)}_{(n)} = m\, {\rm I}
\psi^{(1,1,0)}_n (k^2)\ , \nn \\
&1^{++}:&\quad   \widehat \Psi^{(1,1)}_{(n)\,\mu} =
\sqrt{\frac{3}{2s}}\,i\,
\varepsilon_{\gamma P k \mu} \psi^{(1,1,1)}_n (k^2)\ ,
\nn \\
&2^{++}:&\quad   \widehat \Psi^{(1,2)}_{(n)\,\mu_1\mu_2} =
\sqrt{\frac34}
\left [k_{\mu_1}\gamma^\perp_{\mu_2} +k_{\mu_2}\gamma^\perp_{\mu_1}
-\frac 23 \hat k g^\perp_{\mu_1 \mu_2}\right ]
\psi^{(1,1,2)}_n (k^2)+ \nn \\
&&~~+ \frac{5}{\sqrt 2}\left [k_{\mu_1}k_{\mu_2}\hat k -
\frac 15 k^2(g^\perp_{\mu_1\mu_2}\hat k +\gamma^\perp_{\mu_1}k_{\mu_2}
+k_{\mu_1}\gamma^\perp_{\mu_1})\right ]
\psi^{(1,3,2)}_n (k^2),\nn \\
&1^{+-}:&\quad   \widehat \Psi^{(0,1)}_{(n)\,\mu} =
\sqrt 3\,i\gamma_5 k_{\mu}
\psi^{(0,1,1)}_n (k^2)\ .
\label{wf-1}
\eea
Here, the short notations are used $k^\perp \equiv k=(k_1-k_2)/2$
($(kP)=0$ at quark masses equal to each other) and
$\varepsilon_{\gamma P k \mu}
\equiv \varepsilon_{\nu_1\nu_2\nu_3\mu}\gamma_{\nu_1}P_{\nu_2}k_{\nu_3}$
as well as the equality $\hat k =m{\rm I}$
(recall again that in the spectral integral technique the constituents
are mass-on-shell).

The $1^{--}$ and $2^{++}$ states are defined by two wave
functions with different
$L$; correspondingly, there are two wave functions for these two
levels.

The $1^{--}$ state is characterized by two angular
momenta, $L=0$ and $L=2$. The states in
(\ref{wf-1}) are not pure with respect to  $L$, for the wave function
$\gamma^\perp_\mu\psi^{(1,0,1)}(k^2)$ is a mixture of $S$- and
$D$-waves. The pure $S$ state is given by the following operator (e.g., see
\cite{deut,YFtensor}):
\be \Gamma_\mu\ =\
\gamma^\perp_\mu-\frac{k_\mu}{2m+\sqrt{s}} \label{wf-2}\ .
\ee
In the nonrelativistic limit, the operators $\Gamma_\mu$ and
$\gamma^\perp_\mu$ coincide:
$\left[\Gamma_\mu\simeq\gamma^\perp_\mu\right ]_{nonrel}$.
Likewise, the pure $P$ state of the system $2^{++}$ is defined by the
following operator
\cite{YFtensor}:
\be T_{\mu_1\mu_2}\ =\
\sqrt{\frac34}\left[k_{\mu_1}\Gamma_{\mu_2}+
k_{\mu_2}\Gamma_{\mu_1}-\frac23(k\Gamma)g^\perp_{\mu_1\mu_2}\right]\ .
\label{wf-3}
\ee
Note that for heavy quarks the use of pure operators
(\ref{wf-2}), (\ref{wf-3}) results in the insignificant modification of
the wave function representation, for the $\gamma^\perp_\mu$
operator leads to a small admixture of the $D$-wave  and, vice versa,
the $\frac3{\sqrt 2}\left [k_\mu\hat k-\frac13
k^2\gamma^{\perp}_\mu\right ]$ operator results in a state with
dominant $D$-wave, with a small admixture of the $S$-wave. Similarly,
for the state $2^{++}$ the operator
$\sqrt{\frac34}\left[k_{\mu_1}\gamma^\perp_{\mu_2} +
k_{\mu_2}\gamma^\perp_{\mu_1}
-\frac 23\hat k g^\perp_{\mu_1 \mu_2}\right ]$
leads to the dominant
$P$-wave, while  the operator $\frac5{\sqrt 2}
\left [k_{\mu_1}k_{\mu_2}\hat
k - \frac 15 k^2(g^\perp_{\mu_1\mu_2}\hat k
+\gamma^\perp_{\mu_1}k_{\mu_2} +k_{\mu_1}\gamma^\perp_{\mu_1})\right ]$
gives us the dominant $F$-wave.

Let us note that dealing with the operators of pure states,
like (\ref{wf-2}) and (\ref{wf-3}), would not facilitate the fitting
procedure, for real states are the mixture of different waves
($S$, $D$ for $1^{--}$ and $P$,  $F$ for  $2^{++}$).

\subsection{Radiative transitions $(Q\bar Q)_{in}\to\gamma
(Q\bar Q)_{out}$}

Here, we list the formulas for the amplitudes and partial widths
of the radiative transitions $(Q\bar Q)_{in}\to\gamma (Q\bar Q)_{out}$
used in the fit. The technique for the calculation of the
transition amplitudes $(Q\bar Q)_{in}\to\gamma (Q\bar Q)_{out}$
was developed in \cite{physrev,epja,YFscalar,YFtensor}.
In \cite{radtransitions}, the transition form factors were transformed
to the form convenient for the fitting procedure, and we present them in
this form below.

\subsubsection{Transitions of the vector ($1^{--}$) and
pseudoscalar ($0^{-+}$) mesons}

Here, we present the formulas for the radiative  transitions of the
vector ($V$) and pseudoscalar ($P$) mesons. The partial widths for the
decays $V\to \gamma P$ and $P\to \gamma V$ read:
\bea M_V \Gamma_{V\to
\gamma P}&=& \frac{1}{24}\alpha\frac{(M_V^2-M_P^2)^3}{M_V^2}\left |
F_{V\to \gamma P} \right |^2,\nonumber \\
M_P \Gamma_{P\to \gamma V}&=&
\frac{1}{8}\alpha\frac{(M_P^2-M_V^2)^3}{M_P^2}\left |
F_{V\to \gamma P} \right |^2\, ,
\eea
where $\alpha=e^2/4\pi =1/137$.
The transition form factor  is expressed in terms of the $Q\bar Q$
wave functions $\psi^{(0,0,0)}_n (k^2)$ and $\psi^{(1,0,1)}_n (k^2)$,
see (\ref{wf-1}). However, in line with
\cite{epja,YFscalar,YFtensor},
we change the notations as follows:
\be \psi^{(0,0,0)}_n
(k^2)\equiv\psi_P(s), \quad \psi^{(1,0,1)}_n(k^2) \equiv\psi_{V(0)}(s),
\quad \psi^{(1,2,1)}_n(k^2) \equiv\psi_{V(2)}(s) \ .
\ee
In terms of
$\psi_V(s)$ and $\psi_P(s)$, the form factors are written as:
\bea
F_{V(0)\to \gamma P}= && Z_{V\to \gamma P}
\frac{m}{4\pi}\int\limits_{4m^2}^{\infty}
\frac{ds}{\pi} \psi_{V(0)}(s)\psi_P(s)
\ln\frac{\sqrt{s}+\sqrt{s-4m^2}}{\sqrt{s}-\sqrt{s-4m^2}}\ ,
\\
F_{V(2)\to \gamma P}= && Z_{V\to \gamma P}
\frac{m}{32\sqrt 2\, \pi}\int\limits_{4m^2}^{\infty}
\frac{ds}{\pi} \psi_{V(2)}(s)\psi_P(s)\times
\\ \nonumber
&&\times
\left[(2m^2+s)\ln\frac{\sqrt{s}+\sqrt{s-4m^2}}{\sqrt{s}-\sqrt{s-4m^2}}
-3\sqrt{s(s-4m^2)}\right]\ ,
\eea
 where $Z_{V\to \gamma P}$ is the charge factor:
\be
\label{20}
Z_{V\to \gamma P}=2e_Q, \qquad e_c=\frac 23\, ,\quad e_b=-\frac 13 \ .
\ee
The total transition form factor reads:
\be
F_{V\to \gamma P}=F_{V(0)\to \gamma P}\, +\, F_{V(2)\to \gamma P}\ .
\label{20a}
\ee
Normalization conditions  are determined by
Eq. (\ref{norm_3}), they can be written, after the integration over the
angle variables, as the integrals over $s$:
 \be
\int\!\frac{d^3k}{(2\pi^3)k_0} \to
 \int\limits_{4m^2}^{\infty}
 \frac{ds}{16\pi^2}
\sqrt{\frac{s-4m^2}{s}}\,.
 \ee
Normalization conditions for vector state determined by
Eq. (\ref{norm_3}) read:
\bea
\label{21-V}
1 &=& W_{00}[V]+W_{02}[V]+W_{22}[V], \\
W_{00}[V]&=&\frac13\int\limits_{4m^2}^{\infty}
\frac{ds}{16\pi^2}\ \psi_{V(0)}^2(s)\ 4\left(s+2m^2\right)
\sqrt{\frac{s-4m^2}{s}}\ ,
\\ \nn
W_{02}[V]&=&\frac{\sqrt 2}{3}\int\limits_{4m^2}^{\infty}
\frac{ds}{16\pi^2}\ \psi_{V(0)}(s)\psi_{V(2)}(s)\ (s-4m^2)^2  \,
\sqrt{\frac{s-4m^2}{s}}\ ,
\\ \nn
W_{22}[V]&=&\frac23\int\limits_{4m^2}^{\infty}
\frac{ds}{16\pi^2}\ \psi_{V(2)}^2(s)\ \frac{(8m^2+s)(s-4m^2)^2}{16}
\sqrt{\frac{s-4m^2}{s}}\,.
\eea
Normalization condition for $\psi_P(s)$ in the $s$-integral
representation is:
\bea
1&=&\int\limits_{4m^2}^{\infty}
\frac{ds}{16\pi^2}\ \psi_P^2(s)\ 2s\
\sqrt{\frac{s-4m^2}{s}}\ .
\label{21}
\eea

\subsubsection{Transitions of the vector ($1^{--}$) and
scalar ($0^{++}$) mesons}

The partial widths of  the vector
($V$) and scalar ($S$) meson  decays,
$V\to \gamma S$ and $S\to \gamma V$, read:
\bea
M_V \Gamma_{V\to \gamma S}&=&
\frac{1}{6}\alpha\frac{M_V^2-M_S^2}{M_V^2}\left |
F_{V\to \gamma S} \right |^2,\nonumber \\
M_S \Gamma_{S\to \gamma V}&=&
\frac{1}{2}\alpha\frac{M_S^2-M_V^2}{M_S^2}\left |
F_{V\to \gamma S} \right |^2\, .  \label{22}
\eea
The transition form factor $F_{V\to \gamma S}$ is expressed in terms of
the wave functions $\psi^{(1,0,1)}_n(k^2)$,
$\psi^{(1,2,1)}_n(k^2)$ and $\psi^{(1,1,0)}_n(k^2)$.
Changing the notation in line with,
\cite{epja,YFscalar,YFtensor}
\be
\psi^{(1,1,0)}_n(k^2) \equiv \psi_S(s)\, ,   \label{23}
\ee
we have:
\bea
F_{V(0)\to \gamma S}& = &Z_{V\to \gamma S}\, \frac{m^2}{4\pi}
\int\limits_{4m^2}^{\infty}
\frac{ds}{\pi} \psi_{V(0)}(s)\psi_S(s) I_{V\to \gamma S}(s)
  ,\nn
\\
F_{V(2)\to \gamma S} & =&Z_{V\to \gamma S}\, \frac{\sqrt{2}m^2}{16\pi}
\int\limits_{4m^2}^{\infty}
\frac{ds}{\pi} \psi_{V(2)}(s)\psi_S(s) (-s+4m^2)I_{V\to \gamma S}(s) ,
\nn \\
I_{V\to \gamma S}(s) & =&\sqrt{s(s-4m^2)}-2m^2
\ln\frac{\sqrt{s}+\sqrt{s-4m^2}}{\sqrt{s}-\sqrt{s-4m^2}}\, .
\label{24a}
 \eea

The total form factor is equal to
\bea
F_{V\to \gamma S}&=&F_{V(0)\to \gamma S}\, +\, F_{V(2)\to \gamma S}\, .
\label{24}
\eea
The charge factor of the transition $V\to\gamma S$ for heavy quarks
coincides with that for $V\to\gamma P$:
$Z_{V\to \gamma S}=Z_{V\to \gamma P}$, see Eq. (\ref{20}).
Normalization condition for
 $\psi_S(s)$ reads:
\bea
1&=&\int\limits_{4m^2}^{\infty}
\frac{ds}{16\pi^2}\ \psi_S^2(s)\ 2m^2\left(s-4m^2\right)
\sqrt{\frac{s-4m^2}{s}}
\ .   \label{25}
\eea
Working with the transition amplitude $V\to \gamma S$, one faces the
ambiguity in writing the spin operator that is due to the
existence of the nilpotent spin operator component --- this problem is
discussed in detail in \cite{maxim,kalash}.

\subsubsection{Transitions of the tensor ($2^{++}$)  and
vector ($1^{--}$)  mesons}

Three independent spin operators determine the
transition amplitude $T\to \gamma V$ and, correspondingly, we have
three
form factors \cite{radtransitions}.
In terms of the form factors $F^{(i)}_{T\to\gamma V}$
($i=1,2,3$), the partial
widths for the decays $T\to \gamma V$ and  $V\to \gamma T$ read:
\bea
m_T \Gamma_{T\to\gamma V}&=&
\frac {\alpha}{20}\frac{m_T^2-m_V^2}{m_T^2}\;
\left[
 z^\perp_{11}\;(F^{(1)}_{T\to\gamma V})^2
+z^\perp_{22}\;(F^{(2)}_{T\to\gamma V})^2
+z^\perp_{33}\;(F^{(3)}_{T\to\gamma V})^2
\right],
\nn
\\
m_V \Gamma_{V\to\gamma T}&=&
\frac {\alpha}{12}\frac{m_V^2-m_T^2}{m_V^2}\;
\left[
z^\perp_{11}\;(F^{(1)}_{T\to\gamma V})^2+
z^\perp_{22}\;(F^{(2)}_{T\to\gamma V})^2+
z^\perp_{33}\;(F^{(3)}_{T\to\gamma V})^2
\right]\ .
\eea
Here,
\bea
z^\perp_{11}&=&\frac
{3 M_T^4+34  M_T^2M_V^2+3  M_V^4}{12 M_T^2M_V^2  }   \, ,
\nn
\\
z^\perp_{22}&=&9\, \frac{ M_T^4+10  M_T^2M_V^2 + M_V^4}
{3 M_T^4+34  M_T^2M_V^2 +3 M_V^4  }  \, ,
\nn
\\
z^\perp_{33}&=&\frac 92\,\frac
{(M_T^2+M_V^2)^2}{M_T^4+10  M_T^2M_V^2+  M_V^4}  \, .
\eea
The quark--antiquark $2^{++}$ state is determined by two components
of the wave function
with the dominant $P$- and $F$-waves:
\be
\psi^{(1,1,2)}_n (k^2) \equiv  \psi_{T(1)}(s)  , \quad
\psi^{(1,3,2)}_n (k^2) \equiv  \psi_{T(3)}(s).
\ee
With these notations,  the form factors $F_{T(L)\to\gamma V(L')}^{(i)}$
for $i=1,2,3$
read:
\bea
F_{T(L)\to\gamma V(L')}^{(i)}&=&Z_{T(L)\to\gamma V(L')}
\int\limits_{4m^2}^\infty \frac{ds}{16\pi^2}
S^{(i)}_{T(L)\to\gamma V(L')}(s) \psi_{T}(s)\psi_{V}(s) \, .
\eea
Here,
\bea
S^{(1)}_{T(1)\to\gamma V(0)}(s)&=&
-\frac{\sqrt 3}{5} (8m^2+3s) I^{(1)}_{T\to\gamma V}(s)\ ,
\nn \\
S^{(2)}_{T(1)\to\gamma V(0)}(s)&=&
\frac23S^{(3)}_{T(1)\to\gamma V(0)}(s)
=-\frac{2}{3\sqrt{3}} I^{(2)}_{T\to\gamma V}(s)\ ,
\nn \\
S^{(1)}_{T(1)\to\gamma V(2)}(s)&=&
-\frac{\sqrt{6}}{40} (16m^2-3s)(4m^2-s) I^{(1)}_{T\to\gamma V}(s)\ ,
\nn \\
S^{(2)}_{T(1)\to\gamma V(2)}(s)&=&
\frac23S^{(3)}_{T(1)\to\gamma V(2)}(s)
= -\frac{\sqrt{2}}{12\sqrt{3}}(8m^2+s) I^{(2)}_{T\to\gamma V}(s)\ ,
\nn \\
S^{(1)}_{T(3)\to\gamma V(0)}(s)&=&
-\frac{3\sqrt{2}}{20} (4m^2-s)^2 I^{(1)}_{T\to\gamma V}(s)\ ,
\nn \\
S^{(2)}_{T(3)\to\gamma V(0)}(s)&=&
\frac23S^{(3)}_{T(3)\to\gamma V(0)}(s)
= -\frac{\sqrt{2}}{18}(6m^2+s) I^{(2)}_{T\to\gamma V}(s)\ ,
\nn \\
S^{(1)}_{T(3)\to\gamma V(2)}(s)&=&
-\frac{3}{80} (4m^2-s)^2(8m^2+s) I^{(1)}_{T\to\gamma V}(s)\ ,
\nn \\
S^{(2)}_{T(3)\to\gamma V(2)}(s)&=&
\frac 23S^{(3)}_{T(3)\to\gamma V(2)}(s)
= -\frac{1}{72} (16m^2-3s)(4m^2-s) I^{(2)}_{T\to\gamma V}(s)\ ,
\eea
where
\bea   \label{31b}
I^{(1)}_{T\to\gamma V}(s)&=&
2m^2\ln\frac{\sqrt{ s}+\sqrt {s-4m^2}}
                  {\sqrt{ s}-\sqrt {s-4m^2}}-\sqrt {s(s-4m^2)},
                  \nn \\
I^{(2)}_{T\to\gamma V}(s)&=&
                  m^2(m^2+s)
\ln\frac{\sqrt{ s}+\sqrt {s-4m^2}}{\sqrt{ s}-\sqrt {s-4m^2}}
-\frac 1{12}\sqrt {s(s-4m^2)}(s+26m^2).
\eea

The total form factor is a sum over four terms:
\be
F^{(i)}_{T\to\gamma V}=\sum\limits_{L,L'} F_{T(L)\to\gamma V(L')}^{(i)} .
\label{31a}
\ee
Normalization condition for the tensor meson wave function
is written in the $s$-integral representation as follows:
\bea
\label{32}
1&=&W_{11}[T]+W_{13}[T]+W_{33}[T],\\
W_{11}[T]&=& \frac15\int \limits_{4m^2}^\infty \frac {ds}{16\pi^2}
\; \psi_{T(1)}^2(s)\; \frac 12 (8m^2+3s)(s-4m^2)\rhosq ,
\\ \nn
W_{13}[T]&=& \frac15\int \limits_{4m^2}^\infty \frac {ds}{16\pi^2} \;
\psi_{T(1)}(s)\psi_{T(3)}(s)\;
\frac {\sqrt 3}{2\sqrt 2} (s-4m^2)^3\, \rhosq ,
\\ \nn
W_{33}[T]&=& \frac15\int \limits_{4m^2}^\infty \frac {ds}{16\pi^2} \;
\psi_{T(3)}^2(s)\; \frac 1{16} (6m^2+s)(s-4m^2)^3\, \rhosq \ .
\eea

\subsubsection{Transitions of the pseudovector ($1^{++}$)  and
vector ($1^{--}$)  mesons}

Here we present the formulas for the radiative  transitions of the
vector ($V$) and pseudovector ($A$) mesons. The partial widths for the
decays $A\to \gamma V$ and  $V\to \gamma A$ are determined by the form
factor $F_{A\to\gamma V}$ as follows \cite{radtransitions}:
\bea
\label{33}
m_{A} \Gamma_{A\to\gamma V} &=&\frac
{\alpha}{12}\frac{m_{A}^2-m_{V}^2}{m_{A}^2}\; z^\perp _{A
V}\;F_{A\to\gamma V}^2\, , \nn \\
 m_{V} \Gamma_{V\to\gamma A} &=&\frac
{\alpha}{12}\frac{m_{V}^2-m_{A}^2}{m_{V}^2}\; z^\perp _{VA}
\;F_{A\to\gamma V}^2\, , \eea
where
\be
\label{34}
 z^\perp _{A V}
=-\frac{M^4_A+6M^2_AM^2_V+M^4_V}{2M^2_V} \, ,\quad z^\perp _{VA}
=-\frac{M^4_A+6M^2_AM^2_V+M^4_V}{2M^2_A} \, .
\ee
Changing notation
\be
\label{35} \psi^{(1,1,1)}_n (k^2) = \psi_{A}(s)\ ,
\ee
we write:
\bea
F_{A\to\gamma V(L)}
&=&Z_{A\to\gamma V(L)}
\int \limits_{4m^2}^\infty \frac{ds}{16\pi^2}S_{A\to\gamma V(L)}(s)
\psi_{A}(s)\psi_{V(L)}(s)\ ,
\nn \\
 S_{A\to\gamma V(0)}(s)&=&-\sqrt{\frac32} I^{(1)}_{A\to\gamma V}(s),
\nn \\
 S_{A\to\gamma V(2)}(s)&=& \frac{\sqrt 3}{8}(4m^2-s)I^{(1)}_{A\to\gamma V}(s),
 \label{36}
\eea
where
\bea
I^{(1)}_{A\to\gamma V}(s)&=&
\sqrt{s}\left(2m^2\ln\frac{\sqrt{ s}+\sqrt {s-4m^2}}
             {\sqrt{ s}-\sqrt {s-4m^2}}-\sqrt {s(s-4m^2)}\right).
\eea
The normalization condition reads:
\bea
\label{37}
1= \frac12\int \limits_{4m^2}^\infty \frac {ds}{16\pi^2} \;
\psi_{A}^2(s)\; s(s-4m^2)\rhosq.
\eea

\subsection{Radiative transitions $e^+e^-\to V(Q\bar Q)$
and $Q\bar Q$-$meson\to \gamma\gamma$}

For the consideration
of the radiative transition processes $e^+e^-\to V(Q\bar Q)$
and $\gamma\gamma \to Q\bar Q$-$meson$, it is convenient  to
introduce  the quark--antiquark components of the photon wave function
\cite{epja,PR-g,YF-g}.
 The quark components of photon is tightly
related to the determination  of the quark wave functions of
vector mesons.

The introduction of the quark--antiquark photon wave function may be
illustrated by the two-photon meson decay.
Dealing with the time-ordered processes, that is necessary in the
dispersion relation  or light-cone variable approaches, the
$Q\bar Q$-$meson\to \gamma\gamma$ decay should be treated as a two-step
reaction: the  emission of photon by the quark (Fig. 1a) or antiquark
(Fig. 1b) and a subsequent annihilation $Q\bar Q\to \gamma$. The
triangle diagram cuttings related to these two subprocesses are shown
in Fig. 1c, thus leading  to the representation of the triangle diagram
in terms of the double dispersion integral. In the diagram 1c, on the
left from the first cutting,  there is the transition vertex of
$quarkonium \to Q\bar Q$: we denote this vertex as $G_{Q\bar Q}(s)$.
This vertex determines the wave function of the initial $Q\bar
Q$-meson:
\be \frac{G_{Q\bar Q}(s)}{s-M^2}=\psi_{Q\bar Q}(s)\, .
\label{38} \ee
In the previous subsection, this wave function was denoted
for different $J^{PC}$ as $\psi_{V}(s) $, $\psi_{P}(s)$, $\psi_{S}(s)$,
and so on.

 Likewise, the right-hand cut in Fig. 1c describes the
transition $Q\bar Q\to \gamma$ and provides us with the factor
\be
 \label{39}
\frac1{s'}\, e_Q\ ,
\ee
where $s'$ is the invariant energy square of quarks in the final state
and $e_Q$ is the charge of the $Q$-quark. But when we deal with the
transition $Q\bar Q\to \gamma$, the interaction of quarks should be
necessarily taken into consideration.

The quarks may interact both in initial (Fig. 1d) and final (Fig.
1e) states. In fact, the interaction of quarks in the initial states
has been accounted for in  (\ref{38}), because the vertex functions
$G_{Q\bar Q}$ (or wave functions $\psi_{P}$, $\psi_{S}$, and so on)
are the solutions of the spectral integral equation --- this
equation is shown diagrammatically in Fig. 1f.
As concerns the quark interaction in the final
state, it should be specially taken into account, in addition to
the pointlike interaction (\ref{39}). The diagram shown in Fig. 1g
stands for the quark interaction in the transition $Q\bar Q\to \gamma$,
and we approximate it with the sum of pole terms of the vector meson
($\Upsilon$'s or $\psi$'s
 in the cases of $b\bar b$ or $c\bar c$
systems), see Fig. 1h. Accordingly, the
factor related to the right-hand cut of Fig. 1c is written as follows:
\be
 \label{40}
\frac{G_{\gamma\to Q\bar Q}(s')}{s'}\, e_Q\ ,
\ee
where the vertex function $G_{\gamma\to Q\bar Q}(s')$ at $s'\sim 4m_Q^2$
is the superposition of vertices  of the $V(n)$-mesons (see
Fig. 1h):
\be
 \label{41}
G_{\gamma\to Q\bar Q }(s)\simeq \sum_n C_n G_{V(n)}(s)\ , \quad
s\sim 4m_Q^2\ .
\ee
Here, $n$ is the radial quantum number of $V$-meson and $C_n$'s are
the coefficients which should be determined in the fit.

At large $s$, the vertex $Q\bar Q\to \gamma$ is a pointlike one:
\be
 \label{42}
G_{\gamma\to Q\bar Q}(s)\simeq 1 \qquad {\rm at}\quad s>s_0\ .
\ee
The parameter $s_0$ can be determined from the data on  the
$e^+e^-$-annihilation into hadrons: it defines the energy range where
the ratio $R(s)=\sigma(e^+e^- \to hadrons)/
\sigma(e^+e^- \to \mu^+\mu^-)$ reaches a constant-behavior regime
above the threshold of the production of heavy mesons. The data
\cite{PDG} give us $s_0 \sim (10-15)$ GeV$^2$ for the $c\bar c$
component
 and $s_0 \sim (100-150)$ GeV$^2$ for the $b\bar b$ one.

Therefore, to describe the transition $Q\bar Q \to \gamma$ we may
introduce the characteristics which, similarly to (\ref{38}), can be
called the $Q\bar Q$-component of the photon wave function:
\be
 \label{43}
\frac{G_{\gamma\to Q\bar Q}(s)}{s-q^2}=\Psi_{\gamma(q^2)\to Q\bar Q}(s)
\ .
\ee
Here $q$ is the photon four-momentum. Let us emphasize that such a wave
function is determined at $s\ga 4m^2_Q$.

There are more reactions which are promptly determined by the photon
wave function. These are the transitions $e^+e^- \to \Upsilon$
and $e^+e^- \to \psi$, see Fig.
2: here the loop diagram is defined by the convolution of the vector
meson wave function and $G_{\gamma\to Q\bar Q}$.

The transition $\gamma\to Q\bar Q$  is determined by two spin
structures, $ \gamma_\alpha$ and
$\frac32\left [k_\alpha\hat k-\frac13 k^2\gamma^{\perp}_\alpha\right ]$
(see Eq.(\ref{wf-1})) and, correspondingly, by two vertices.

\be
\label{44}
\gamma_\alpha G^{(S)}_{\gamma\to Q\bar Q}(s) \ , \qquad
\gamma_\xi X^{(2)}_{\xi\alpha} G^{(D)}_{\gamma\to Q\bar Q}(s)
\ee
It means
that we take into account the normal quark--photon interaction,
$\gamma_\alpha$, as well as the contribution of the anomalous magnetic
moment.

For the vertex function of the transition $\gamma\to Q\bar Q$ we use
the following fitting  formula:
\bea
\label{44G}
&&G^{(S)}_{\gamma\to Q\bar Q}(s)=
\sum\limits_{n=1}^6 C_{nS} G_{V(nS)}(s)+
\frac 1{1+\exp(-\beta_\gamma (s-s_0))}\ ,
\\ \nn
&&G^{(D)}_{\gamma\to Q\bar Q}(s)=
\sum\limits_{n=1}^6 C_{nD} G_{V(nD)}(s)\ ,
\nonumber
\eea
where
$G_{V (nS)}(s) = \psi_{n}^{(101)}(s) (s-M^2_{V(nS)})$ and
$G_{V (nD)}(s) = \psi_{n}^{(121)}(s) (s-M^2_{V(nD)})$. The parameters
$C_{nS}$, $C_{nD}$,  $\beta_\gamma$ and $s_0$  are determined in the fit.

\subsubsection{Decay $V\to e^+e^-$}

The transition amplitudes $V\to e^+e^-$ were calculated in
\cite{YF-g,YF-g-cc}. The partial width for the decay $V\to
e^+e^-$ reads:
\bea
\label{45}
\Gamma(V\to e^+e^-)=
\frac{\pi\alpha^2}{M_{V}^5}
\sqrt{\frac{M_{V}^2-4\mu_e^2}{M_{V}^2}}
\left(\frac 83 \mu_e^2+\frac 43 M_{V}^2\right)
\left |F_{V\to e^+e^-}\right |^2,
\eea
where
$\mu_e$ is the electron mass, $M_{V}$ is the measured
vector quarkonium mass, and the form factor $F_{V\to e^+e^-}$ is
determined by the process of Fig. 2.
The quark loops in Fig. 2 are different for the $S$- and $D$-wave
$Q\bar Q$
states, so we have two
 transition amplitudes $F_{V(n,L)\to e^+e^-}$, with
$L=0$ and $L=2$:
 \bea
F_{V\to e^+e^-} = F_{V(0)\to e^+e^-}\, +\, F_{V(2)\to e^+e^-} \ ,
\eea
where
\bea
\label{46}
F_{V(0)\to e^+e^-}=Z^{V\to e^+e^-}_{Q\bar Q} \sqrt{N_c}
\int\limits_{4m^2}^\infty \frac {ds}{16\pi^2}\psi_{V(0)}(s)
G^{(S)}_{\gamma\to Q\bar Q}(s) \left(\frac 83 m^2+\frac 43 s\right)\rhosq\ ,
\eea
$$
\nonumber
F_{V(2)\to e^+e^-}=Z^{V\to e^+e^-}_{Q\bar Q} \sqrt{N_c}
\int\limits_{4m^2}^\infty \frac {ds}{16\pi^2}\psi_{V(2)}(s)
G^{(D)}_{\gamma\to Q\bar Q}(s)\frac{(s-4m^2)^2}{6}\rhosq .
$$
At $L=0,2$, the wave functions $\psi_{V(L)}(s)$  are normalized
according to (\ref{21-V}); here the charge factors are
$Z^{V\to e^+e^-}_{b\bar b} =-1/3$ and $Z^{V\to e^+e^-}_{c\bar c} =2/3$.

\subsubsection{Decay $P\to \gamma\gamma$}
The two-photon decays of pseudoscalar $q\bar q$ mesons $(L=0)$ were
studied in \cite{epja,PR-g}.
Partial width for the decay $P\to\gamma\gamma$ reads:
\bea
\label{48}
\Gamma(P\to\gamma\gamma)=\frac\pi4
\alpha^2 M_{P}^3 \left |F_{P\to\gamma\gamma}\right |^2.
\eea
The transition amplitude is determined by the processes of Figs.
1$a,b$, it reads: \bea \label{49}
F_{P\to\gamma\gamma}=Z^{\gamma\gamma}_{Q\bar Q} \sqrt{N_c}\,  m
\int\limits_{4m^2}^\infty \frac {ds}{2\pi^2}\psi_{P}(s)
\Psi_{\gamma \to Q\bar Q}(s)
\ln\frac{\sqrt{s}+\sqrt{s-4m^2}}{\sqrt{s}-\sqrt{s-4m^2}}\ .
\eea
Recall that $\Psi_{\gamma \to Q\bar Q}(s)=G_{\gamma \to Q\bar Q} (s)/s
$. To underline the existence of  decays with different radial excited
states, we introduce the index $n$ in (\ref{49}). Normalization of
$\psi_{P}(s)$ is given by (\ref{21}), and
\be \label{z-gg}
Z^{\gamma\gamma}_{b\bar b} =\frac 29, \qquad
Z^{\gamma\gamma}_{c\bar c} =\frac 89 .
\ee

\subsubsection{Decay $S\to \gamma\gamma$}

The two-photon decay of the $0^{++}q\bar q$ mesons $(L=1)$ was
considered in \cite{epja,YFscalar}. The partial width of the decay
$S(n)\to\gamma\gamma$ ($n$ is radial quantum number) reads:
\bea
\label{50}
\Gamma(S\to\gamma\gamma)=
\frac{\pi\alpha^2}{M_{S}}
\left |F_{S(n)\to\gamma\gamma}\right |^2\ ,
\eea
with the quark transition amplitude (Figs. 1a,b) equal to:
\bea
\label{51}
F_{S(n)\to\gamma\gamma}=Z^{\gamma\gamma}_{Q\bar Q} \sqrt{N_c}\, m^2
\int\limits_{4m^2}^\infty \frac {ds}{4\pi^2}\psi_{S(n)}(s)
\Psi_{\gamma \to Q\bar Q}(s)\times
\\ \nonumber
\times \left(\sqrt{s(s-4m^2)}
-2m^2\ln\frac{\sqrt{s}+\sqrt{s-4m^2}}{\sqrt{s}-\sqrt{s-4m^2}}
\right).
\eea
Normalization of $\psi_{S(n)}(s)$ is given by (\ref{25}).

\subsubsection{Decay $T\to \gamma\gamma$}

The two-photon tensor meson decay amplitude was calculated in
\cite{epja,YFtensor}.
The partial width for the decay process $T\to\gamma\gamma$
is defined by two transition amplitudes with the helicities $H=0,2$:
\bea
\label{52}
\Gamma(T\to\gamma\gamma)=\frac
45\frac{\pi\alpha^2}{M_{T}} \left [\frac 16
\left |F_{T\to\gamma\gamma}^{(0)}\right|^2 +
\left|F_{T\to\gamma\gamma}^{(2)}\right|^2\right].
\eea
Taking into account
the $P$- and $F$-wave quark--antiquark component, we write the form
factors $F_{T\to\gamma\gamma}^{(H)}$ as
\bea
\label{52a}
F_{T\to\gamma\gamma}^{(H)} = F_{T(1)\to\gamma\gamma}^{(H)} \, +
\, F_{T(3)\to\gamma\gamma}^{(H)} \ ,
\eea
where
\bea
\label{53}
F_{T(L)\to\gamma\gamma}^{(H)}=Z^{\gamma\gamma}_{Q\bar Q} \sqrt{N_c}
\int\limits_{4m^2}^\infty \frac {ds}{16\pi^2}\psi_{T(L)}(s)
\Psi_{\gamma \to Q\bar Q}(s)S_{T(L)\to\gamma\gamma}^{(H)} (s)\ ,
\eea
with the following spin factors:
\bea
S_{T(1)\to\gamma\gamma}^{(0)}(s)&=&
-\frac{4}{\sqrt 3}\sqrt{s\left(s-4m^2\right)}
\left(12m^2+s\right)
+\frac{8m^2}{\sqrt 3}\left(4m^2+3s\right)
\ln\frac{s+\sqrt{s\left(s-4m^2\right)}}
        {s-\sqrt{s\left(s-4m^2\right)}} \ ,
\nn \\
S_{T(3)\to\gamma\gamma}^{(0)}(s)&=&
-\frac{2\sqrt{2s\left(s-4m^2\right)}}{5}
\left(72m^4+8m^2s+s^2\right)+
\\ \nonumber
&&+\frac{12\sqrt 2}{5}m^2\left(8m^4+4m^2s+s^2\right)
\ln\frac{s+\sqrt{s\left(s-4m^2\right)}}{s-\sqrt{s\left(s-4m^2\right)}}
\ ,
\label{54}
\eea
and
\bea
S_{T(1)\to\gamma\gamma}^{(2)}(s)&=&\frac{8\sqrt{s\left(s-4m^2\right)}}
{3\sqrt 3}
\left(5m^2+s\right)
-\frac{8m^2}{\sqrt 3}\left(2m^2+s\right)
\ln\frac{s+\sqrt{s\left(s-4m^2\right)}}
        {s-\sqrt{s\left(s-4m^2\right)}}\ , \nn \\
S_{T(3)\to\gamma\gamma}^{(2)}(s)&=&
\frac{2\sqrt{2s\left(s-4m^2\right)}}{15}
\left(30m^4-4m^2s+s^2\right)-
\\ \nonumber
&&-\frac{2\sqrt 2}{5}m^2\left(12m^4-2m^2s+s^2\right)
\ln\frac{s+\sqrt{s\left(s-4m^2\right)}}{s-\sqrt{s\left(s-4m^2\right)}}
\ .
\label{55}
\eea
Normalization of $\psi_{T{(1)}}(s)$ and $\psi_{T{(3)}}(s)$ is
determined by (\ref{32}).

\section{Bottomonium states found from
spectral integral equation and radiative transitions}

The quarkonium wave functions are fitted in the following form:
\bea
\label{21-1}
\psi^{(S,L,J)}_{(n)}(k^2)=e^{-\beta k^2}\sum\limits_{i=1}^9 c_i(S,L,J;n)
k^{i-1}\, ,
\eea
where we re-denoted $k^2\equiv {\bf k}^2$;
recall that $s=4m^2+4{\bf k}^2$. The
fitting parameter $\beta$ is of the order of $0.5-1.5 $ GeV$^{-2}$ and
may be different for different flavor sectors. We put here $\beta =1.2$
GeV$^{-2}$

The data in the $b\bar b $ sector can be described
by two types of the $t$-channel exchanges,
scalar and vector ones:
$ {\rm I}\otimes {\rm I},\quad
\gamma_{\mu\ }\otimes \gamma_{\mu}$.   The
addition of the pseudoscalar exchanges like
$\gamma_5\otimes \gamma_5$ does not improve the fit.

Here, we present
three variants of fit: with instantaneous forces (solution $I(b\bar b)$),
retarded interactions (solution $R(b\bar b)$) and that with the universal
"confinement potential" -- instantaneous
interaction with nearly the same
parameters $a$ and $b$ in (\ref{2-2-14})
for all the quark sectors: $b\bar b $, $c\bar c $ and $q\bar q $
 (solution $U(b\bar b)$) .

For the $b\bar b$ sector, the parameters for  scalar and vector
exchange interactions (${\rm I}\otimes {\rm I}$ and
$\gamma_{\mu}\otimes \gamma_{\mu}$, see Section 2.2) are as follows
(all values are in GeV):
\be
\begin{tabular}{ccccccccc}
Interaction & Wave & $a$  &  $b$ & $c$ & $\mu_c$ & $d$ & $\mu_d$ \\
\hline
$({\rm I} \otimes {\rm I})$ &
\begin{tabular}{c}
$I(b\bar b)$ \\  $U(b\bar b)$ \\ $R(b\bar b)$
\end{tabular}
&
\begin{tabular}{c}
                     -0.151 \\  0.911 \\ -0.680
\end{tabular}
&
\begin{tabular}{c}
                      0.160 \\  0.150 \\  0.130
\end{tabular}
&
\begin{tabular}{c}
                      0.506 \\ -0.377 \\  1.322
\end{tabular}
&
\begin{tabular}{c}
                      0.201 \\  0.401 \\  0.201
\end{tabular}
&
\begin{tabular}{c}
                     -0.250 \\ -0.201 \\ -0.228
\end{tabular}
&
\begin{tabular}{c}
                      0.201 \\  0.401 \\  0.401
\end{tabular}
\\ \hline
$(\gamma_{\mu} \otimes \gamma_{\mu})$&
\begin{tabular}{c}
$I(b\bar b)$ \\  $U(b\bar b)$ \\  $R(b\bar b)$
\end{tabular}
&
\begin{tabular}{c}
                     -0.812 \\  1.178 \\ -1.620
\end{tabular}
&
\begin{tabular}{c}
                      0.000 \\ -0.150 \\ -0.005
\end{tabular}
&
\begin{tabular}{c}
                      0.867 \\ -1.356 \\  1.821
\end{tabular}
&
\begin{tabular}{c}
                      0.401 \\  0.201 \\  0.201
\end{tabular}
&
\begin{tabular}{c}
                      0.300 \\  0.500 \\  0.311
\end{tabular}
&
\begin{tabular}{c}
                      0.001 \\  0.001 \\  0.001
\end{tabular}
\\ \hline
\end{tabular}
\ee

\subsection{Masses of the $b\bar b$ states}

The fitting procedure prefers for the constituent
$b$-quark the mass value  $m_b=4.5$ GeV. This value looks
quite reasonable if we take into account
that mass difference of the constituent and QCD quarks is of the order
of $200-350$ MeV and
the QCD estimates \cite{m-QCD} give the constraint $4.0\leq m_b(QCD)
\leq 4.5$ GeV.

The masses of  $b\bar b$ states for $n=1,2,3,4,5,6$
(experimental values and those obtained
in the fit) are given below, in (\ref{79}) -- (\ref{84}). The bold
numbers stand for the masses, which are included in the fitting
procedure. In parentheses we show the dominant wave for $b\bar b$
state ($S$ or $D$ for $1^{--}$ and $P$ or $F$ for $2^{++}$). The
right-hand side columns show the mean square radii of bottomonia (in
GeV$^{-2}$ for solution $I(b\bar b)$.

We have the following masses (in GeV) for $1^{--}$ states:
\begin{equation} \label{79}
\begin{tabular}{lllllll}
State          &          Data  & $I(b\bar b)$ & $U(b\bar b)$ & $R(b\bar b)$ &  $R^2_I$  \\
$\Upsilon(1S)$ & {\bf  9.460}   &  9.392 $(S)$ &  9.382 $(S)$ &  9.448 $(S)$ &  0.342   \\
$\Upsilon(2S)$ & {\bf 10.023}   & 10.029 $(S)$ & 10.027 $(S)$ & 10.023 $(S)$ &  1.632   \\
$\Upsilon(1D)$ &      10.150    & 10.159 $(D)$ & 10.158 $(D)$ & 10.105 $(D)$ &  0.342   \\
$\Upsilon(3S)$ & {\bf 10.355}   & 10.368 $(S)$ & 10.365 $(S)$ & 10.362 $(S)$ &  3.794   \\
$\Upsilon(2D)$ &      10.450    & 10.439 $(D)$ & 10.436 $(D)$ & 10.156 $(D)$ &  1.632   \\
$\Upsilon(4S)$ & {\bf 10.580}   & 10.615 $(S)$ & 10.634 $(S)$ & 10.628 $(S)$ &  6.504   \\
$\Upsilon(3D)$ &      10.700    & 10.661 $(D)$ & 10.677 $(D)$ & 10.430 $(D)$ &  3.794   \\
$\Upsilon(5S)$ & {\bf 10.865}   & 10.819 $(S)$ & 10.872 $(S)$ & 10.851 $(S)$ &  9.793   \\
$\Upsilon(4D)$ &      10.950    & 10.852 $(D)$ & 10.898 $(D)$ & 10.669 $(D)$ &  6.504   \\
$\Upsilon(6S)$ & {\bf 11.020}   & 11.019 $(S)$ & 11.084 $(S)$ & 11.040 $(S)$ & 11.990   \\
$\Upsilon(5D)$ & ---            & 11.023 $(D)$ & 11.109 $(D)$ & 10.892 $(D)$ &  9.793   \\
$\Upsilon(6D)$ & ---            & 11.214 $(D)$ & 11.303 $(D)$ & 11.085 $(D)$ & 11.990 , \\
\end{tabular}
\end{equation}
 for $0^{-+}$ states:
\begin{equation}  \label{80}
\begin{tabular}{llllll}
State & Data & $I(b\bar b)$ & $U(b\bar b)$ & $R(b\bar b)$ & $R^2_I$ \\
$\eta_{b}(1S)$ & {\bf 9.300} &  9.334 &  9.322 &  9.393 &  0.922 \\
$\eta_{b}(2S)$ &    ---      & 10.006 & 10.011 & 10.007 &  2.782 \\
$\eta_{b}(3S)$ &    ---      & 10.344 & 10.355 & 10.352 &  5.781 \\
$\eta_{b}(4S)$ &    ---      & 10.557 & 10.626 & 10.621 & 18.839 \\
$\eta_{b}(5S)$ &    ---      & 10.636 & 10.864 & 10.846 & 13.699 \\
$\eta_{b}(6S)$ &    ---      & 10.837 & 11.079 & 11.037 & 11.668 ,
 \end{tabular}
\end{equation}
 for $0^{++}$ states:
\begin{equation} \label{81}
\begin{tabular}{llllll}
State & Data & $I(b\bar b)$ & $U(b\bar b)$ & $R(b\bar b)$ & $R^2_I$ \\
$\chi_{b0}(1P)$ & {\bf  9.859} &  9.852 &  9.862 &  9.851 &  0.847 \\
$\chi_{b0}(2P)$ & {\bf 10.232} & 10.241 & 10.236 & 10.227 &  2.632 \\
$\chi_{b0}(3P)$ &  ---         & 10.509 & 10.517 & 10.510 &  5.161 \\
$\chi_{b0}(4P)$ &  ---         & 10.726 & 10.759 & 10.745 &  8.053 \\
$\chi_{b0}(5P)$ &  ---         & 10.884 & 10.983 & 10.952 & 12.437 \\
$\chi_{b0}(6P)$ &  ---         & 10.947 & 11.185 & 11.084 & 19.969 ,
\end{tabular}
\end{equation}
 for $1^{++}$ states:
\begin{equation} \label{82}
\begin{tabular}{llllll}
State & Data & $I(b\bar b)$ & $U(b\bar b)$ & $R(b\bar b)$ & $R^2_I$ \\
$\chi_{b1}(1P)$ & {\bf  9.892} &  9.884 &  9.895 &  9.890 &  0.915 \\
$\chi_{b1}(2P)$ & {\bf 10.255} & 10.257 & 10.252 & 10.249 &  2.777 \\
$\chi_{b1}(3P)$ &  ---         & 10.516 & 10.528 & 10.526 &  5.814 \\
$\chi_{b1}(4P)$ &  ---         & 10.697 & 10.767 & 10.762 & 18.944 \\
$\chi_{b1}(5P)$ &  ---         & 10.759 & 10.989 & 10.970 & 13.544 \\
$\chi_{b1}(6P)$ &  ---         & 10.920 & 11.191 & 11.199 & 11.702 ,
\end{tabular}
\end{equation}
 for $2^{++}$ states:
\begin{equation}\label{83}
\begin{tabular}{llllll}
State           &           Data & $I(b\bar b)$ & $U(b\bar b)$ & $R(b\bar b)$ & $R^2_I$ \\
$\chi_{b2}(1P)$ & {\bf  9.912} &  9.909 $(P)$ &  9.911 $(P)$ &  9.923 $(P)$ &   0.956 \\
$\chi_{b2}(2P)$ & {\bf 10.268} & 10.270 $(P)$ & 10.262 $(P)$ & 10.268 $(P)$ &   2.782 \\
$\chi_{b2}(1F)$ & ---          & 10.365 $(F)$ & 10.347 $(F)$ & 10.346 $(F)$ &   0.956 \\
$\chi_{b2}(3P)$ & ---          & 10.528 $(P)$ & 10.535 $(P)$ & 10.538 $(P)$ &   5.361 \\
$\chi_{b2}(2F)$ & ---          & 10.594 $(F)$ & 10.592 $(F)$ & 10.592 $(F)$ &   2.782 \\
$\chi_{b2}(4P)$ & ---          & 10.738 $(P)$ & 10.773 $(P)$ & 10.768 $(P)$ &   8.573 \\
$\chi_{b2}(3P)$ & ---          & 10.788 $(P)$ & 10.813 $(F)$ & 10.807 $(F)$ &   5.361 \\
$\chi_{b2}(5F)$ & ---          & 10.846 $(F)$ & 10.994 $(P)$ & 10.972 $(P)$ &  18.995 \\
$\chi_{b2}(4P)$ & ---          & 10.963 $(P)$ & 11.020 $(F)$ & 11.006 $(F)$ &   8.573 \\
$\chi_{b2}(6F)$ & ---          & 10.931 $(F)$ & 11.196 $(P)$ & 11.150 $(P)$ &  13.978 \\
$\chi_{b2}(5F)$ & ---          & 11.124 $(F)$ & 11.221 $(F)$ & 11.187 $(F)$ &  18.995 \\
$\chi_{b2}(6F)$ & ---          & 11.326 $(F)$ & 11.411 $(F)$ & 11.380 $(F)$ &  13.978 ,
\end{tabular}
\end{equation}
and  for $1^{+-}$ states, $h_b$$(1^{+-})$:
\begin{equation} \label{84}
\begin{tabular}{llllll}
State     & Data & $I(b\bar b)$ & $U(b\bar b)$ & $R(b\bar b)$ & $R^2_I$ \\
$h_b(1S)$ & ---  &  9.889 &  9.902 &  9.896 &  0.922 \\
$h_b(2S)$ & ---  & 10.259 & 10.255 & 10.253 &  2.782 \\
$h_b(3S)$ & ---  & 10.518 & 10.530 & 10.529 &  5.781 \\
$h_b(4S)$ & ---  & 10.700 & 10.768 & 10.764 & 18.839 \\
$h_b(5S)$ & ---  & 10.759 & 10.990 & 10.971 & 13.699 \\
$h_b(6S)$ & ---  & 10.921 & 11.192 & 11.200 & 11.668 .
\end{tabular}
\end{equation}

The wave functions for solution $I(b\bar b)$ are presented in Appendix
II: we show the wave functions in the $k$-representation   and give
coefficients $c_i$, $\beta$ for Eq. (\ref{21-1}). For $1^{--}$ and
$2^{++}$ states, we give $W_{LL'}$: one can see that the admixture
of  second components is small.

\subsection{Radiative decays $(b\bar b)_{in}\to\gamma (b\bar b)_{out}$}

Figure 3 shows  the radiative transitions which are included into the
fitting procedure.

The fit gives us the following values for radiative decay of
$\Upsilon$-mesons (partial widths in keV):
\begin{equation}
\begin{tabular}{lllllll}
Process & Data & $I(b\bar b)$ & $U(b\bar b)$ & $R(b\bar b)$ &  $\cite{Resag}$ & $\cite{Kuhn}$ \\
$\Upsilon (1S) \to \gamma\eta_{b0}(1S)$ &   ---       &   0.0092 &  0.0100 &  0.0079 & --- & ---    \\
$\Upsilon (2S) \to \gamma\eta_{b0}(1S)$ &   ---       &   0.0025 &  0.0015 &  0.0008 & --- & ---    \\
$\Upsilon (2S) \to \gamma\eta_{b0}(2S)$ &   ---       &   0.0006 &  0.0002 &  0.0002 & --- & ---    \\
$\Upsilon (2S) \to \gamma\chi_{b0}(1P)$ & 1.7$\pm$0.2 &   1.1023 &  1.0669 &  0.9611 & 1.62 & 1.41  \\
$\Upsilon (2S) \to \gamma\chi_{b1}(1P)$ & 3.0$\pm$0.5 &   2.7288 &  2.3675 &  2.1268 & 2.55 & 2.27  \\
$\Upsilon (2S) \to \gamma\chi_{b2}(1P)$ & 3.1$\pm$0.5 &   2.9100 &  2.6674 &  1.9367 & 2.51 & 2.24  \\
$\Upsilon (3S) \to \gamma\eta_{b0}(1S)$ &   ---       &   0.0016 &  0.0007 &  0.0005 & --- & ---    \\
$\Upsilon (3S) \to \gamma\eta_{b0}(2S)$ &   ---       &   0.0013 &  0.0000 &  0.0002 & --- & ---    \\
$\Upsilon (3S) \to \gamma\eta_{b0}(3S)$ &   ---       &   0.0006 &  0.0001 &  0.0001 & --- & ---    \\
$\Upsilon (3S) \to \gamma\chi_{b0}(2P)$ & 1.4$\pm$0.2 &   1.1191 &  1.3746 &  1.1253 & 1.77 & ---   \\
$\Upsilon (3S) \to \gamma\chi_{b1}(2P)$ & 3.0$\pm$0.5 &   3.4368 &  4.0831 &  3.1279 & 2.88 & ---   \\
$\Upsilon (3S) \to \gamma\chi_{b2}(2P)$ & 3.0$\pm$0.5 &   3.7266 &  4.7438 &  3.1686 & 3.14 & ---   \\
\label{rd-1}
\end{tabular}
\end{equation}
 For the
illustration, in (\ref{rd-1}) we present the results of Refs.
\cite{Resag,Kuhn}.

The radiative decay of $\chi_{bJ}$ is not included into  fitting
procedure. We have the following predictions for the partial widths
(in keV):
\begin{equation}
\begin{tabular}{lllll}
Process & Data & $I(b\bar b)$ & $U(b\bar b)$ & $R(b\bar b)$ \\
$\chi_{b0}(1P)\to\gamma\Upsilon (1S)$   &
               $<\Gamma_{tot}(\chi_{b0}(1P))\cdot 6\cdot 10^{-2}$
               & 43.49 & 52.79 & 34.47 \\
$\chi_{b1}(1P)\to\gamma\Upsilon (1S)$   &
               $\Gamma_{tot}(\chi_{b1}(1P))\cdot (35\pm 8)\cdot
               10^{-2}$  & 53.31 & 63.77 & 42.62 \\
$\chi_{b2}(1P)\to\gamma\Upsilon (1S)$   &
               $\Gamma_{tot}(\chi_{b2}(1P))\cdot (22\pm 4)\cdot
               10^{-2}$  & 47.61 & 56.15 & 38.10 \\
$\chi_{b0}(2P)\to\gamma\Upsilon (1S)$   &
               $\Gamma_{tot}(\chi_{b0}(2P))\cdot (0.9\pm 0.6)\cdot 10^{-2}$
                                                                          &    7.49 &  9.25 &  7.08 \\
$\chi_{b0}(2P)\to\gamma\Upsilon (2S)$   &
               $\Gamma_{tot}(\chi_{b0}(2P))\cdot (4.6\pm 2.1)\cdot 10^{-2}$
                                                                          &   11.94 & 15.88 & 11.06 \\
$\chi_{b1}(2P)\to\gamma\Upsilon (1S)$   &
               $\Gamma_{tot}(\chi_{b1}(2P))\cdot (8.5\pm 1.3)\cdot 10^{-2}$
                                                                          &   12.76 & 16.85 & 11.69 \\
$\chi_{b1}(2P)\to\gamma\Upsilon (2S)$   &
               $\Gamma_{tot}(\chi_{b1}(2P))\cdot (21.0\pm 4.0)\cdot 10^{-2}$
                                                                          &   12.92 & 14.40 & 16.01 \\
$\chi_{b2}(2P)\to\gamma\Upsilon (1S)$   &
               $\Gamma_{tot}(\chi_{b2}(2P))\cdot (7.1\pm 1.0)\cdot 10^{-2}$
                                                                          &   18.11 & 20.58 & 12.37 \\
$\chi_{b2}(2P)\to\gamma\Upsilon (2S)$   &
               $\Gamma_{tot}(\chi_{b2}(2P))\cdot (16.2\pm 2.4)\cdot 10^{-2}$
                                                      & 16.14 & 18.25 & 14.41 \\
$\chi_{b0}(3P)\to\gamma\Upsilon (1S)$   &   ---       &  2.69 &  3.56 &  2.84 \\
$\chi_{b1}(3P)\to\gamma\Upsilon (1S)$   &   ---       &  4.41 &  7.06 &  5.43 \\
$\chi_{b2}(3P)\to\gamma\Upsilon (1S)$   &   ---       &  5.60 &  8.11 &  6.01 \\
$\chi_{b0}(3P)\to\gamma\Upsilon (2S)$   &   ---       &  2.05 &  1.86 &  2.04 \\
$\chi_{b1}(3P)\to\gamma\Upsilon (2S)$   &   ---       &  3.38 &  3.88 &  4.09 \\
$\chi_{b2}(3P)\to\gamma\Upsilon (2S)$   &   ---       &  4.37 &  4.59 &  4.57 \\
$\chi_{b0}(3P)\to\gamma\Upsilon (3S)$   &   ---       &  7.59 & 10.37 &  7.19 \\
$\chi_{b1}(3P)\to\gamma\Upsilon (3S)$   &   ---       & 10.74 & 15.85 & 11.33 \\
$\chi_{b2}(3P)\to\gamma\Upsilon (3S)$   &   ---       &  9.90 & 13.81 &  9.87 \\
\end{tabular}
\label{rd-2}
\end{equation}
The total widths  $\Gamma_{tot}(\chi_{bJ}(1P)$ and
$\Gamma_{tot}(\chi_{bJ}(2P)$, with $J=0,1,2$, have not been measured
yet.

The calculations performed on the basis of (\ref{rd-2}) give us the
following estimates for total widths:
\bea
\Gamma_{tot}(\chi_{b0}(1P)) &   <& 730\ {\rm keV},          \\ \nn
\Gamma_{tot}(\chi_{b1}(1P)) & \simeq & 120 - 200\ {\rm keV}, \\ \nn
\Gamma_{tot}(\chi_{b2}(1P)) & \simeq & 180 - 270\ {\rm keV}, \\ \nn
\Gamma_{tot}(\chi_{b0}(2P)) & \simeq & 180 - 480\ {\rm keV},\\ \nn
\Gamma_{tot}(\chi_{b1}(2P)) & \simeq & 50 - 80\ {\rm keV},  \\ \nn
\Gamma_{tot}(\chi_{b2}(2P)) & \simeq & 70 - 120\ {\rm keV}.
\eea
The
fit gives us the following values for partial widths of radiative
decays of $\eta_{b0}$-mesons:
\begin{equation}
\begin{tabular}{ccccc}
Process & Data & $I(b\bar b)$ & $U(b\bar b)$ & $R(b\bar b)$ \\
$\eta_{b0}(2S) \to \gamma\Upsilon(1S)$ & --- & 0.21 & 0.20 & 0.12 \\
$\eta_{b0}(3S) \to \gamma\Upsilon(1S)$ & --- & 0.27 & 0.18 & 0.11 \\
$\eta_{b0}(3S) \to \gamma\Upsilon(2S)$ & --- & 0.05 & 0.02 & 0.01 \\
\end{tabular} \end{equation}

\subsection{The $b\bar b$ component of the photon
wave function}

Fitting to the reactions with the $\gamma\to b\bar
b$ transitions, we  determined the parameters
$C_n,\beta_\gamma,s_0$ defined in (\ref{44G}) for
$G^{S,D}_{\gamma\to b\bar b}(s)$.
For solutions $R(b\bar b)$, $I(b\bar b)$ and $U(b\bar b)$, they
are as follows (in GeV):
 \begin{equation}
\begin{tabular}{lll}
$I(b\bar b)$         & $U(b\bar b)$         & $R(b\bar b)$      \\
$C_{1S}$ = -0.263    & $C_{1S}$ = -0.800   & $C_{1S}$ = -0.220 \\
$C_{2S}$ = -0.296    & $C_{2S}$ = -0.303   & $C_{2S}$ =  0.092 \\
$C_{3S}$ =  0.057    & $C_{3S}$ =  0.074   & $C_{3S}$ = -0.038 \\
$C_{4S}$ =  0.298    & $C_{4S}$ =  0.197   & $C_{4S}$ = -0.027 \\
$C_{5S}$ = -2.000    & $C_{5S}$ = -0.781   & $C_{5S}$ =  0.262 \\
$C_{6S}$ =  1.093    & $C_{6S}$ =  2.000   & $C_{6S}$ = -0.441 \\
$C_{1D}$ = -0.554    & $C_{1D}$ = -0.328   & $C_{1D}$ =  0.168 \\
$C_{2D}$ = -0.284    & $C_{2D}$ =  0.233   & $C_{2D}$ =  0.109 \\
$b_\gamma$ = 2.85    & $b_\gamma$ = 2.85    & $b_\gamma$ = 2.85 \\
$s_0$ =    18.79     & $s_0$     = 18.79    & $s_0$ =    18.79
\end{tabular}
\label{37a}
\end{equation}
Experimental values of partial widths included into fitting procedure
as an input
together with those obtained in the fitting procedure are shown below:
\begin{equation}
\begin{tabular}{cccccc}
Process & Data          & $I(b\bar b)$ & $U(b\bar b)$ & $R(b\bar b)$ & \cite{Gonz}
  \\
$\Upsilon(1S)\to e^+e^-$ & $1.314\pm$0.029  & 1.314 & 1.313 & 1.314  & 1.01 \\
$\Upsilon(2S)\to e^+e^-$ & $0.576\pm$0.024  & 0.576 & 0.575 & 0.576  & 0.35 \\
$\Upsilon(3S)\to e^+e^-$ & $0.476\pm$0.076  & 0.476 & 0.476 & 0.476  & 0.25 \\
$\Upsilon(4S)\to e^+e^-$ & $0.248\pm$0.031  & 0.248 & 0.248 & 0.248  & 0.22 \\
$\Upsilon(5S)\to e^+e^-$ & $0.31\pm$0.07    & 0.310 & 0.310 & 0.310  & 0.18 \\
$\Upsilon(6S)\to e^+e^-$ & $0.130\pm$0.03   & 0.130 & 0.130 & 0.130  & 0.14
\end{tabular}
\end{equation}
The last column in (90) demonstrates the results of \cite{Gonz}.

 The vertices $G_{\gamma\to b\bar b}(s)$, which represent our solutions
and given in
(\ref{37a}), are shown
shown in Fig 4. The data and predictions for the  two-photon
partial widths $\eta_{b0}\to\gamma\gamma$, $\chi_{b0}\to\gamma\gamma$,
$\chi_{b2}\to\gamma\gamma$ are as follows:
\begin{equation}
\begin{tabular}{lllllllllll}
Process & Data & $I(b\bar b)$ & $U(b\bar b)$ & $R(b\bar b)$ &
\cite{G} & \cite{M} & \cite{Gupta2} & \cite{Sch} & \cite{Huang} &
\cite{Barnes} \\\\ $\eta_{b0}(1S)\to\gamma\gamma$ & --- & 1.554 & 1.851
& 1.537 & 0.35   & 0.22   & 0.46   & 0.46   & 0.45 & 0.17 \\
$\eta_{b0}(2S)\to\gamma\gamma$ & --- & 1.928 & 2.296 & 1.906 & 0.11   & ---    & 0.20   & 0.21   & 0.13 & ---  \\
$\eta_{b0}(3S)\to\gamma\gamma$ & --- & 2.139 & 2.547 & 2.115 & 0.10   & 0.084  & ---    & ---    & ---  & ---  \\
$\chi_{b0}(1P)\to\gamma\gamma$ & --- & 0.024 & 0.029 & 0.021 & 0.038  & 0.024  & 0.080  & 0.043  & ---  & --- \\
$\chi_{b0}(2P)\to\gamma\gamma$ & --- & 0.023 & 0.028 & 0.021 & 0.029  & 0.026  & ---    & ---    & ---  & --- \\
$\chi_{b0}(3P)\to\gamma\gamma$ & --- & 0.023 & 0.027 & 0.020 & ---    & ---    & ---    & ---    & ---  & --- \\
$\chi_{b2}(1P)\to\gamma\gamma$ & --- & 0.016 & 0.020 & 0.013 & 0.0080 & 0.0056 & 0.0080 & 0.0074 & ---  & --- \\
$\chi_{b2}(2P)\to\gamma\gamma$ & --- & 0.015 & 0.020 & 0.013 & 0.0060 & 0.0068 & ---    & ---    & ---  & --- \\
$\chi_{b2}(3P)\to\gamma\gamma$ & --- & 0.015 & 0.019 & 0.012 & ---    & ---    & ---    & ---    & ---  & ---
\end{tabular}
\end{equation}
The last six columns represent the calculation results of [40--45].

\subsection{Potentials in the solution $I(b\bar b)$ }

To be illustrative, let us demonstrate the interaction in one of the
solutions, say, $I(b\bar b)$, in the language of  potentials.

In the solution $I(b\bar b)$, the confinement potential is due to
scalar exchanges only, $V^{(S)}_{conf}(r)=-0.151+0.160r$ (GeV),
it is shown in Fig. 5a. For the illustration, we demonstrate  also
the $0^{-+}$-levels created by this confinement potential alone.
However, the scalar exchange has the short-range component
$V^{(S)}_{short}(r)=0.506\exp(-0.2r)-0.250/r/exp(-0.2r)$ (GeV), see
Fig. 5b, which pushes  the levels higher. The potential related to the
$t$-channel vector exchanges,
$V^{(V)}(r)=0.812-0.867\exp(-0.4r)-0.300/r$ (GeV), does not contain the
increasing part ($\sim r$), see Fig. 5c. The last term may be
interpreted as a one-gluon exchange, with $\alpha_s=0.300\cdot 3/4
\simeq 0.23$. Also in Fig. 5c,  we demonstrate for the illustration the
$0^{-+}$-levels, which would be created by vector exchange forces only.

In the solution $U(b\bar b)$, the vector-exchange forces contain
the one-gluon exchange term:
$V^{(V)}_{short}(r)=1.355\exp(-0.5r)-0.500/r$ (GeV) which corresponds
to $\alpha_s \simeq 0.38$.

\section{Conclusion}

In the framework of the method, which is in fact a variant of the
dispersion relation approach, we have performed the description of the
bottomonium spectra: the $b\bar b$-levels and their radiative
transitions such as
$(b\bar b)_{in}\to \gamma+(b\bar b)_{out}$ ,
 $e^+e^- \to V(b\bar b)$  and $b\bar b$-$meson\to \gamma  \gamma $.
Using quark--antiquark interaction as an input, we have obtained several
variants of a reasonably good fit of the data. The ambiguities in a
reconstruction of the soft region $b\bar b$ interaction underline the
problem we face: a scarcity of the radiative decay data. To restore
 the $b\bar b$ interaction, one needs much more data, in particular,
on the two-photon reactions: $ \gamma  \gamma\to b\bar b$-$meson$,
including the bottomonium production by virtual photons in $ \gamma
\gamma^*$ and $ \gamma^* \gamma^*$ collisions.

We pay a considerable attention to the presentation of the $b\bar b$
wave functions. The matter is that a solution may represent
rather similarly the state levels, though with different
wave functions. Therefore, we think that solution should be
characterized by two characteristics, that is, the position of the
level and its wave function.

\section*{Acknowledgments}

We thank A.V. Anisovich, Y.I. Azimov, G.S. Danilov, I.T. Dyatlov,
L.N. Lipatov, V.Y. Petrov, H.R. Petry and M.G. Ryskin
for useful discussions.

This work was supported by the Russian Foundation for Basic Research,
project no. 04-02-17091.

\section{Appendix I. The structure of  pseudoscalar,
 scalar and vector exchanges}

The loop diagram, that includes the interaction, is given by the
expression as follows:
\be Sp[Q^{(S,L,J)}_{\mu_1\ldots\mu_J}(m+\hat k_1)O_I(m+\hat
k'_1) Q^{(S,L,J)}_{\nu_1\ldots\nu_J}(m-\hat k'_2)O_I(m-\hat k_2)]=
V^{(S,L,J)}_I (-1)^J O^{\mu_1\ldots\mu_J}_{\nu_1\ldots\nu_J}\,,
\ee
where $k_1,k_2$ are the momenta of particles before the interaction,
$k'_1,k'_2$ are their momenta after the interaction, and the operators
$O_I$'s are given by  (\ref{bs4}).

For the singlet ($S=0$) states in the case of scalar, pseudoscalar
and vector exchanges, we obtain:
\bea
&V^{(0,J,J)}_I&=\sqrt{ss'}
\big (4z\kappa-4m^2-\sqrt{ss'}\big )
\kappa^J P_J(z)\ ,
\nonumber \\
&V^{(0,J,J)}_{\gamma_5}&=\sqrt{ss'}
\big (4z\kappa+4m^2-\sqrt{ss'}\big )
\kappa^J P_J(z)\ ,
\nonumber \\
&V^{(0,J,J)}_{\gamma_\mu}&=\sqrt{ss'}
\big (4\sqrt{ss'}-8m^2\big )
\kappa^J P_J(z)\ .
\eea
Here, $P_J(z)$ are Legendre polynomials depending on the angle between
final and initial particles and
\be
\kappa=|{\bf k}||{\bf k}'|\,.
\label{kappa}
\ee
Near the threshold,
the pseudoscalar interaction has a higher order factors $\kappa$
$|{\bf k}||{\bf k}'|$ than in the case of scalar and vector interactions
thus playing a minor role for mesons consisted of  heavy quarks.
The scalar and vector interactions in the lowest $|{\bf k}||{\bf k}'|$
order are equal to each other in absolute value but have  opposite sign.

To obtain the expressions for triplet states, first, let us calculate
the trace with vertex functions taken as $\gamma_\mu$. Then,
general expression can be obtained by the convolution of the trace
operators:
\bea &&{\rm Sp}[\gamma_\mu (m+\hat k_1)O_I(m+\hat k'_1) \gamma_\nu
(m-\hat k'_2)O_I(m-\hat k_2)]= (a^I_1+z\kappa\, a^I_2)\,
g^\perp_{\mu\nu}+\! a^I_3 k^\perp_\mu k^\perp_\nu\!+ \nonumber \\
&&a^I_4 k'^\perp_\mu k'^\perp_\nu\!+ (a^I_5+z\kappa\,a^I_6)\,
k^\perp_\mu k'^\perp_\nu\! + a^I_7(k^\perp_\mu
k'^\perp_\nu\!-\!k'^\perp_\mu k^\perp_\nu)\ . \label{gg_2}
\eea
The
coefficients $a_i$ for the scalar, pseudoscalar and vector exchanges are
equal to
\bea
\begin{array}{l|c|c|c}
O_I   &  1                              & \gamma_5                       & \gamma_\mu  \\ \hline
a^I_1 & \sqrt{ss'}(4m^2\!+\!\sqrt{ss'}) & \sqrt{ss'}(4m^2\!-\!\sqrt{ss'})& -2ss' \\
a^I_2 & -4\sqrt{ss'}                    & +4\sqrt{ss'}                   & -8\sqrt{ss'} \\
a^I_3 & +4s'                            & -4s'                           & -8s' \\
a^I_4 & +4s                             & -4s                            & -8s \\
a^I_5 & 4(4m^2\!-\!\sqrt{ss'})          & 4(4m^2\!+\!\sqrt{ss'})         & 8(8m^2\!-\!\sqrt{ss'})\\
a^I_6 & -16                             & +16                            & +32 \\
a^I_7 & +4\sqrt{ss'}                    & -4\sqrt{ss'}                   & +8\sqrt{ss'}     \ .
\end{array}
\label{gg_3}
\eea

Then, for $S=1$ and $L=J$ states we obtain:
\bea
&V^{(1,J,J)}_1&=\sqrt{ss'}\kappa^{J}
\left [
\big (4z\kappa\!-\!4m^2\!-\!\sqrt{ss'}\big )P_J(z)-
\frac{4\kappa}{J+1}(zP_J(z)-P_{J-1}(z))
\right ] \ ,
\nonumber \\
&V^{(1,J,J)}_{\gamma_5}&=\sqrt{ss'}
\kappa^{J}
\left [
\big (\!-4z\kappa\!-\!4m^2\!+\!\sqrt{ss'}\big )P_J(z)+
\frac{4\kappa}{J+1}(z P_J(z)-P_{J-1}(z))
\right ]\ ,
\nonumber \\
&V^{(1,J,J)}_{\gamma_\mu}&=\sqrt{ss'}
\kappa^{J}
\left [
\big (2\sqrt{ss'}\!+\!8z\kappa \big )P_J(z)-
\frac{8\kappa}{J+1}(zP_J(z)-P_{J-1}(z))
\right ]\ .
\label{int_3LJ}
\eea

Likewise,  the states with $L=J\pm 1$ are expressed as
follows:
\bea V^{(1,L,L',J)}_I=\frac{-1}{2J+1} \kappa^{\frac{L+L'}{2}}
\sum\limits_{k=1}^7 a^I_k\,v^{(L,L')}_k  \ .
\eea
We use additional index ($L'$) to describe transitions
between states with $L^+\!=\!J\!+\!1$ and $L^-\!=\!J\!-\!1$.
\bea
\begin{array}{l|l|l|l|l}
 ~ & L^-\!\to\!L^-   &   L^+\!\to\!L^+  &
     L^-\!\to\!L^+   &   L^+\!\to\!L^-   \\ \hline
v^{(L,L')}_1 & (2J\!+\!1)P_{J-1}(z)   &
      (2J\!+\!1)P_{J+1}(z)   &
       0   &
       0   \\
v^{(L,L')}_2 & (2J\!+\!1)z\kappa\,P_{J-1}(z)   &
      (2J\!+\!1)z\kappa\,P_{J+1}(z)   &
       0   &
       0   \\
v^{(L,L')}_3 &-JP_{J-1}(z) |{\bf k}|^2   &
     -(J\!+\!1)P_{J+1}(z) |{\bf k}|^2   &
      \Lambda\,\kappa\, P_{J+1} &
      \Lambda\, P_{J-1}\frac{|{\bf k}|^4}{\kappa} \\
v^{(L,L')}_4 &-J\,P_{J-1}(z) |{\bf k}'|^2   &
     -(J\!+\!1)P_{J+1}(z) |{\bf k}'|^2   &
      \Lambda\, P_{J-1}(z)\frac{|{\bf k}'|^4}{\kappa}   &
      \Lambda\, \kappa\,P_{J+1}(z)   \\
v^{(L,L')}_5 &-J\kappa\, P_J(z)  &
     -(J\!+\!1)\kappa\, P_J(z)  &
      \Lambda  P_J(z) |{\bf k}'|^2 &
      \Lambda  P_J(z) |{\bf k}|^2 \\
v^{(L,L')}_6 &-J\,z\kappa^{2} P_J(z) &
     -(J\!+\!1)z\kappa^{2} P_J(z) &
      \Lambda z\kappa P_J(z) |{\bf k}'|^2 &
      \Lambda z\kappa P_J(z) |{\bf k}|^2 \\
v^{(L,L')}_7 & \frac{(2J\!+\!1)(1-J)}{2J-1}\kappa(
       P_J(z)\!-\!P_{J-2}(z))  &
     (2J\!+\!1)\kappa(z P_{J+1}(z)\!-\!P_{J}(z)) &
       0   &
       0  \ . \\
\end{array}
\eea
Here, $\Lambda=\sqrt{J(J+1)}$ and $\kappa$ are defined by
(\ref{kappa}).

\section{Appendix II. Wave functions in the $b\bar b$ sector}

The tables 1--4 demonstrate the coefficient values $c^{(n)}_i$,
which determine the wave functions $\psi^{(S,L,J)}$ according to
(\ref{21-1}). In the tables we also show $W_{LL'}$, which determine
the normalization condition, see (\ref{W}). In Figs. 6--9, we
demonstrate these wave functions.

\begin{table}
\caption{Constants $c_i^{(n)}$ from  (\ref{21-1}) (in GeV) for
the wave functions of the $\Upsilon$-mesons in solution $I(b\bar b)$;
also we present the normalization coefficients $W_{00},W_{02},W_{22}$
given by (12),(13).}

\begin{center}
{\footnotesize
\begin{tabular}{|r|r|r|r|r|r|r|} \hline &
\multicolumn{2}{|c|}{$\Upsilon(1S)$} &
\multicolumn{2}{|c|}{$\Upsilon(2S)$} &
\multicolumn{2}{|c|}{$\Upsilon(3S)$} \\
\hline
 & \multicolumn{2}{|c|}{$W_{00}\qquad W_{02}\qquad W_{22}$} &
   \multicolumn{2}{|c|}{$W_{00}\qquad W_{02}\qquad W_{22}$} &
   \multicolumn{2}{|c|}{$W_{00}\qquad W_{02}\qquad W_{22}$} \\
& \multicolumn{2}{|c|}{1.00050 \, -0.00060 \, 0.00010} &
  \multicolumn{2}{|c|}{1.00101 \, -0.00122 \, 0.00021} &
  \multicolumn{2}{|c|}{1.00115 \, -0.00143 \, 0.00028} \\
\hline
$i$ & $\psi^{(1,0,1)}$ & $\psi^{(1,2,1)}$ & $\psi^{(1,0,1)}$ & $\psi^{(1,2,1)}$  & $\psi^{(1,0,1)}$ & $\psi^{(1,2,1)}$  \\
\hline
 1 &         2.0060 &        -1.9545 &         2.7111 &         1.7177 &        -4.7273 &         1.9830 \\
 2 &       -10.8380 &        13.5543 &         9.8323 &       -12.2344 &         9.5837 &       -12.8048 \\
 3 &        73.0097 &       -36.0805 &       -72.4706 &        32.3653 &       -52.0435 &        33.3246 \\
 4 &      -230.4699 &        47.0494 &       221.2708 &       -41.5308 &       240.2794 &       -42.7176 \\
 5 &       397.0086 &       -30.8840 &      -386.7066 &        26.4284 &      -457.6044 &        26.8052 \\
 6 &      -389.9981 &         7.9660 &       386.9034 &        -6.1016 &       453.3999 &        -5.5935 \\
 7 &       218.1722 &         1.2842 &      -219.7120 &        -1.5700 &      -251.5948 &        -2.0591 \\
 8 &       -64.5122 &        -1.1142 &        65.7237 &         1.0735 &        73.9804 &         1.2288 \\
 9 &         7.8476 &         0.1657 &        -8.0769 &        -0.1519 &        -9.0153 &        -0.1688 \\
\hline
\hline
& \multicolumn{2}{|c|}{$\Upsilon(4S)$} & \multicolumn{2}{|c|}{$\Upsilon(5S)$} &
  \multicolumn{2}{|c|}{$\Upsilon(6S)$} \\
\hline
 & \multicolumn{2}{|c|}{$W_{00}\qquad W_{02}\qquad W_{22}$} &
   \multicolumn{2}{|c|}{$W_{00}\qquad W_{02}\qquad W_{22}$} &
   \multicolumn{2}{|c|}{$W_{00}\qquad W_{02}\qquad W_{22}$} \\
& \multicolumn{2}{|c|}{1.00118 \, -0.00155 \, 0.00037} &
  \multicolumn{2}{|c|}{1.00149 \, -0.00185 \, 0.00036} &
  \multicolumn{2}{|c|}{0.80131 \,  0.03464 \, 0.16405} \\
\hline
$i$ & $\psi^{(1,0,1)}$ & $\psi^{(1,2,1)}$ & $\psi^{(1,0,1)}$ & $\psi^{(1,2,1)}$  & $\psi^{(1,0,1)}$ & $\psi^{(1,2,1)}$  \\
\hline
 1 &        -4.6053 &        -1.9248 &         4.7469 &        -2.3093 &         4.3499 &       -33.8613 \\
 2 &       -12.9581 &        16.3285 &        26.3926 &        11.4831 &        66.6334 &        71.5349 \\
 3 &       122.1101 &       -45.3030 &      -218.0913 &       -23.7269 &      -672.1165 &        95.8832 \\
 4 &      -232.1932 &        57.4674 &       306.9638 &        26.0199 &      2023.7316 &      -345.3417 \\
 5 &       242.2475 &       -34.4863 &       140.7892 &       -15.4406 &     -2851.3726 &       273.7456 \\
 6 &      -211.8979 &         6.2974 &      -597.4634 &         3.9824 &      2118.8719 &       -24.6291 \\
 7 &       139.6440 &         2.9879 &       479.8989 &         0.3477 &      -850.6343 &       -61.8950 \\
 8 &       -50.5914 &        -1.5944 &      -161.4547 &        -0.4002 &       175.5466 &        27.9503 \\
 9 &         7.2976 &         0.2100 &        20.2973 &         0.0570 &       -15.0191 &        -3.6706 \\
\hline
\end{tabular}
}
\end{center}
\end{table}

\begin{table}
\caption{Constants $c_i^{(n)}$ from eq. (\ref{21-1}) (in GeV units) for
the wave functions of $\Upsilon$-mesons in the solution $I(b\bar b)$;
normalization coefficients $W_{00},W_{02},W_{22}$ are
given by (12),(13).}

\begin{center} {\footnotesize
\begin{tabular}{|r|r|r|r|r|r|r|}
\hline
& \multicolumn{2}{|c|}{$\Upsilon(1D)$} & \multicolumn{2}{|c|}{$\Upsilon(2D)$} &
  \multicolumn{2}{|c|}{$\Upsilon(3D)$} \\
\hline
 & \multicolumn{2}{|c|}{$W_{00}\qquad W_{02}\qquad W_{22}$} &
   \multicolumn{2}{|c|}{$W_{00}\qquad W_{02}\qquad W_{22}$} &
   \multicolumn{2}{|c|}{$W_{00}\qquad W_{02}\qquad W_{22}$} \\
& \multicolumn{2}{|c|}{0.00083 \, -0.00214 \, 1.00131} &
  \multicolumn{2}{|c|}{0.00144 \, -0.00360 \, 1.00217} &
  \multicolumn{2}{|c|}{0.00210 \, -0.00499 \, 1.00290} \\
\hline
$i$ & $\psi^{(1,2,1)}$ & $\psi^{(1,0,1)}$ & $\psi^{(1,2,1)}$ & $\psi^{(1,0,1)}$  & $\psi^{(1,2,1)}$ & $\psi^{(1,0,1)}$  \\
\hline
 1 &         0.5316 &         0.0051 &       -14.2711 &        -0.0804 &         6.0089 &         0.0734 \\
 2 &        22.2640 &         0.2961 &        29.5676 &         0.4676 &       104.3726 &         0.4895 \\
 3 &       -61.8266 &        -2.1943 &       -62.5673 &        -3.1827 &      -367.7237 &        -4.4838 \\
 4 &        79.4272 &         7.1306 &       100.0037 &        11.9276 &       446.4882 &        12.5128 \\
 5 &       -50.4830 &       -12.7875 &       -73.1973 &       -21.9636 &      -236.8819 &       -20.8090 \\
 6 &        10.6083 &        12.9455 &        15.1171 &        22.1691 &        31.8558 &        22.5104 \\
 7 &         4.5874 &        -7.4078 &         8.7956 &       -12.6656 &        23.2527 &       -14.3861 \\
 8 &        -2.8530 &         2.2290 &        -5.1628 &         3.8222 &       -10.8663 &         4.7718 \\
 9 &         0.4300 &        -0.2751 &         0.7691 &        -0.4749 &         1.4288 &        -0.6355 \\
\hline
\hline
& \multicolumn{2}{|c|}{$\Upsilon(4D)$} & \multicolumn{2}{|c|}{$\Upsilon(5D)$} &
  \multicolumn{2}{|c|}{$\Upsilon(6D)$} \\
\hline
 & \multicolumn{2}{|c|}{$W_{00}\qquad W_{02}\qquad W_{22}$} &
   \multicolumn{2}{|c|}{$W_{00}\qquad W_{02}\qquad W_{22}$} &
   \multicolumn{2}{|c|}{$W_{00}\qquad W_{02}\qquad W_{22}$} \\
& \multicolumn{2}{|c|}{0.00230 \, -0.00585 \, 1.00355} &
  \multicolumn{2}{|c|}{0.20433 \, -0.04594 \, 0.84162} &
  \multicolumn{2}{|c|}{0.00126 \, -0.00134 \, 1.00008} \\
\hline
$i$ & $\psi^{(1,2,1)}$ & $\psi^{(1,0,1)}$ & $\psi^{(1,2,1)}$ & $\psi^{(1,0,1)}$  & $\psi^{(1,2,1)}$ & $\psi^{(1,0,1)}$  \\
\hline
 1 &       -32.4189 &        -0.0913 &        80.9176 &         2.0418 &      -377.8732 &         0.5165 \\
 2 &       -32.2563 &        -0.3537 &      -211.1041 &        33.2603 &      2121.7911 &        -1.1281 \\
 3 &       385.0197 &         1.9186 &       -54.4316 &      -333.9307 &     -4529.4955 &       -14.6612 \\
 4 &      -592.6611 &         6.0232 &       550.7145 &      1012.1198 &      4646.2765 &        69.2347 \\
 5 &       321.7644 &       -30.8630 &      -476.3406 &     -1440.4555 &     -2245.1955 &      -123.0232 \\
 6 &       -16.3314 &        44.9067 &        41.8114 &      1084.8218 &       251.1282 &       109.8199 \\
 7 &       -45.1952 &       -29.9473 &       113.3274 &      -443.3884 &       204.4454 &       -52.1311 \\
 8 &        15.1386 &         9.5444 &       -50.9272 &        93.7641 &       -81.9575 &        12.5435 \\
 9 &        -1.4251 &        -1.1922 &         6.6630 &        -8.2835 &         9.2592 &        -1.2099 \\
\hline
\end{tabular}
}
\end{center}
\end{table}

\begin{table}
\caption{Constants $c_i^{(n)}$ from (\ref{21-1}) (in GeV) for
the wave functions of $\chi_{b2}$-mesons in the solution $I(b\bar b)$
normalization coefficients $W_{00},W_{02},W_{22}$ are
given by (12),(13).}

\begin{center}
{\footnotesize
\begin{tabular}{|r|r|r|r|r|r|r|}
\hline
& \multicolumn{2}{|c|}{$\chi_{b2}(1P)$} & \multicolumn{2}{|c|}{$\chi_{b2}(2P)$} &
  \multicolumn{2}{|c|}{$\chi_{b2}(3P)$} \\
\hline
 & \multicolumn{2}{|c|}{$W_{11}\qquad W_{13}\qquad W_{33}$} &
   \multicolumn{2}{|c|}{$W_{11}\qquad W_{13}\qquad W_{33}$} &
   \multicolumn{2}{|c|}{$W_{11}\qquad W_{13}\qquad W_{33}$} \\
& \multicolumn{2}{|c|}{1.00075 \, -0.00094 \, 0.00019} &
  \multicolumn{2}{|c|}{1.00128 \, -0.00203 \, 0.00075} &
  \multicolumn{2}{|c|}{0.98981 \, -0.00494 \, 0.01514} \\
\hline
$i$ & $\psi^{(1,1,2)}$ & $\psi^{(1,3,2)}$ & $\psi^{(1,1,2)}$ & $\psi^{(1,3,2)}$  & $\psi^{(1,1,2)}$ & $\psi^{(1,3,2)}$  \\
\hline
 1 &        13.0540 &        -4.1772 &       -34.6278 &        16.0313 &      -257.7504 &       100.2644 \\
 2 &      -114.9374 &        23.4967 &       456.0158 &       -90.5844 &      2844.0283 &      -561.9424 \\
 3 &       487.8876 &       -52.5492 &     -1939.7651 &       201.0440 &    -12019.5471 &      1240.5753 \\
 4 &     -1062.2386 &        60.1441 &      4168.0416 &      -227.9811 &     25892.6834 &     -1399.5333 \\
 5 &      1317.9252 &       -37.5852 &     -5113.9135 &       141.4654 &    -31653.5893 &       863.3132 \\
 6 &      -967.3832 &        12.2496 &      3712.1978 &       -46.2459 &     22808.4240 &      -280.4788 \\
 7 &       414.8491 &        -1.5240 &     -1571.7829 &         6.1338 &     -9564.2325 &        37.0697 \\
 8 &       -95.8539 &        -0.1252 &       357.7836 &         0.2835 &      2152.2305 &         1.6274 \\
 9 &         9.2113 &         0.0356 &       -33.7704 &        -0.1074 &      -200.3714 &        -0.6277 \\
\hline
\hline
& \multicolumn{2}{|c|}{$\chi_{b2}(4P)$} & \multicolumn{2}{|c|}{$\chi_{b2}(5P)$} &
  \multicolumn{2}{|c|}{$\chi_{b2}(6P)$} \\
\hline
 & \multicolumn{2}{|c|}{$W_{11}\qquad W_{13}\qquad W_{33}$} &
   \multicolumn{2}{|c|}{$W_{11}\qquad W_{13}\qquad W_{33}$} &
   \multicolumn{2}{|c|}{$W_{11}\qquad W_{13}\qquad W_{33}$} \\
& \multicolumn{2}{|c|}{1.00071 \, -0.00246 \, 0.00176} &
  \multicolumn{2}{|c|}{1.00246 \, -0.00318 \, 0.00072} &
  \multicolumn{2}{|c|}{1.00311 \, -0.00397 \, 0.00086} \\
\hline
$i$ & $\psi^{(1,1,2)}$ & $\psi^{(1,3,2)}$ & $\psi^{(1,1,2)}$ & $\psi^{(1,3,2)}$  & $\psi^{(1,1,2)}$ & $\psi^{(1,3,2)}$  \\
\hline
 1 &        61.1310 &       -27.9538 &       -21.7721 &         4.7026 &        10.8365 &        -2.3155 \\
 2 &      -804.3179 &       157.6290 &        68.2854 &       -23.9703 &       144.6924 &         5.6067 \\
 3 &      3433.8802 &      -347.3610 &      -194.8198 &        48.5937 &      -929.0568 &         2.8923 \\
 4 &     -7271.1541 &       388.8039 &       630.5268 &       -51.7634 &      1868.8590 &       -20.3080 \\
 5 &      8755.3125 &      -236.5028 &     -1080.4598 &        32.0724 &     -1686.1180 &        23.4272 \\
 6 &     -6242.9707 &        74.8939 &       933.0968 &       -12.0176 &       679.5364 &       -11.5459 \\
 7 &      2595.9632 &        -9.1414 &      -421.0616 &         2.7404 &       -61.8407 &         2.2550 \\
 8 &      -579.2785 &        -0.6539 &        94.9352 &        -0.3698 &       -31.7481 &         0.0334 \\
 9 &        53.3862 &         0.1874 &        -8.3610 &         0.0242 &         6.7593 &        -0.0471 \\
\hline
\hline
& \multicolumn{2}{|c|}{$\chi_{b2}(1F)$} & \multicolumn{2}{|c|}{$\chi_{b2}(2F)$} &
  \multicolumn{2}{|c|}{$\chi_{b2}(3F)$} \\
\hline
 & \multicolumn{2}{|c|}{$W_{11}\qquad W_{13}\qquad W_{33}$} &
   \multicolumn{2}{|c|}{$W_{11}\qquad W_{13}\qquad W_{33}$} &
   \multicolumn{2}{|c|}{$W_{11}\qquad W_{13}\qquad W_{33}$} \\
& \multicolumn{2}{|c|}{0.00058 \, -0.00191 \, 1.00132} &
  \multicolumn{2}{|c|}{0.00097 \, -0.00315 \, 1.00218} &
  \multicolumn{2}{|c|}{0.00124 \, -0.00390 \, 1.00266} \\
\hline
$i$ & $\psi^{(1,3,2)}$ & $\psi^{(1,1,2)}$ & $\psi^{(1,3,2)}$ & $\psi^{(1,1,2)}$  & $\psi^{(1,3,2)}$ & $\psi^{(1,1,2)}$  \\
\hline
 1 &        -3.6248 &        -0.0963 &       -14.7982 &         0.0560 &       -37.1706 &         2.9107 \\
 2 &        -7.2244 &         0.8055 &         0.1033 &        -1.0323 &         9.5966 &       -34.8030 \\
 3 &        19.9292 &        -2.9097 &        31.5118 &         3.4274 &       147.9125 &       151.9065 \\
 4 &       -17.8259 &         5.3091 &       -19.0518 &        -4.1874 &      -177.1377 &      -332.1606 \\
 5 &         5.4336 &        -5.0342 &        -2.9800 &         1.6200 &        36.1365 &       410.3068 \\
 6 &         2.6649 &         2.4892 &         3.3342 &         1.1027 &        48.9104 &      -298.9169 \\
 7 &        -2.9157 &        -0.5206 &         1.0840 &        -1.4679 &       -35.1616 &       126.9788 \\
 8 &         0.9614 &        -0.0152 &        -0.9814 &         0.5902 &         9.1555 &       -29.0017 \\
 9 &        -0.1144 &         0.0160 &         0.1614 &        -0.0846 &        -0.8818 &         2.7474 \\
\hline
\hline
& \multicolumn{2}{|c|}{$\chi_{b2}(4F)$} & \multicolumn{2}{|c|}{$\chi_{b2}(5F)$} &
  \multicolumn{2}{|c|}{$\chi_{b2}(6F)$} \\
\hline
 & \multicolumn{2}{|c|}{$W_{11}\qquad W_{13}\qquad W_{33}$} &
   \multicolumn{2}{|c|}{$W_{11}\qquad W_{13}\qquad W_{33}$} &
   \multicolumn{2}{|c|}{$W_{11}\qquad W_{13}\qquad W_{33}$} \\
& \multicolumn{2}{|c|}{0.00244 \, -0.00497 \, 1.00253} &
  \multicolumn{2}{|c|}{0.00119 \, -0.00465 \, 1.00346} &
  \multicolumn{2}{|c|}{0.20223 \,  0.04827 \, 0.74950} \\
\hline
$i$ & $\psi^{(1,3,2)}$ & $\psi^{(1,1,2)}$ & $\psi^{(1,3,2)}$ & $\psi^{(1,1,2)}$  & $\psi^{(1,3,2)}$ & $\psi^{(1,1,2)}$  \\
\hline
 1 &       137.2096 &        -3.9590 &      -295.5265 &         0.0318 &      -342.8625 &        88.3149 \\
 2 &      -395.5970 &        55.7513 &      1172.2960 &        -8.1102 &      1605.5381 &      -670.5111 \\
 3 &       346.9297 &      -259.4898 &     -1660.4991 &        60.4182 &     -2779.6979 &      1944.5425 \\
 4 &       -64.2683 &       578.6142 &       946.4669 &      -176.4908 &      2084.3230 &     -2837.7766 \\
 5 &       -14.0966 &      -709.0203 &       -65.5667 &       263.0556 &      -374.7669 &      2282.8464 \\
 6 &       -40.3793 &       502.8571 &      -139.7540 &      -217.9552 &      -406.6671 &     -1033.5668 \\
 7 &        38.2866 &      -205.4367 &        44.2943 &       101.2739 &       272.3230 &       257.4060 \\
 8 &       -11.2262 &        44.7930 &        -0.5172 &       -24.6525 &       -64.5095 &       -33.5769 \\
 9 &         1.0954 &        -4.0251 &        -0.8887 &         2.4484 &         5.5083 &         2.1172 \\
\hline
\end{tabular}
}
\end{center}
\end{table}

\begin{table}
\caption{Constants $c^{(n)}_i$ from  (\ref{rd-1}) (in GeV)
for wave functions $\eta_{b0}$, $\chi_{b0}$, $\chi_{b1}$ and , $h_{b1}$
in solution $I(b\bar b)$}
\begin{center} {\footnotesize
\begin{tabular}{|r|r|r|r|r|r|r|}
\hline
$i$ & $\eta_{b0}(1P)$ & $\eta_{b0}(2P)$ & $\eta_{b0}(3P)$ &
      $\eta_{b0}(4P)$ & $\eta_{b0}(5P)$ & $\eta_{b0}(6P)$  \\
\hline
 1 &         4.5338 &        -1.1326 &        11.9074 &       -18.3892 &        29.5281 &        -3.5400 \\
 2 &       -55.4035 &        73.3964 &      -136.8130 &       411.7847 &      -480.8391 &       201.4420 \\
 3 &       329.9544 &      -440.1212 &       798.7926 &     -2495.6136 &      2863.6920 &     -1321.6947 \\
 4 &      -942.3382 &      1240.9035 &     -2311.1680 &      6933.5200 &     -8155.6641 &      3472.8097 \\
 5 &      1474.8959 &     -1923.5277 &      3573.4205 &    -10437.4060 &     12468.2246 &     -4685.6254 \\
 6 &     -1326.7453 &      1712.0637 &     -3132.4041 &      9006.4575 &    -10761.0759 &      3555.8036 \\
 7 &       682.8131 &      -870.3125 &      1560.9990 &     -4427.9932 &      5234.8626 &     -1537.4724 \\
 8 &      -186.4237 &       234.3339 &      -411.2328 &      1148.6087 &     -1334.8254 &       352.9427 \\
 9 &        20.9478 &       -25.9307 &        44.4414 &      -121.6868 &       138.3033 &       -33.1141 \\
\hline
\hline
$i$ & $\chi_{b0}(1P)$ & $\chi_{b0}(2P)$ & $\chi_{b0}(3P)$ &
      $\chi_{b0}(4P)$ & $\chi_{b0}(5P)$ & $\chi_{b0}(6P)$  \\
\hline
 1 &        29.2420 &       -61.6937 &      -219.3775 &       126.2557 &       -42.4376 &        -6.8271  \\
 2 &      -304.0472 &       755.4596 &      2411.7486 &     -1536.0712 &       316.4890 &       347.2500  \\
 3 &      1304.2479 &     -3236.5223 &    -10285.1211 &      6586.8209 &     -1275.2749 &     -1772.0896  \\
 4 &     -2863.9307 &      7047.5098 &     22382.0814 &    -14186.9213 &      2956.9372 &      3636.4974  \\
 5 &      3585.8612 &     -8744.9017 &    -27622.3384 &     17341.7384 &     -3876.4948 &     -3793.2353  \\
 6 &     -2656.7748 &      6413.5855 &     20084.3830 &    -12510.0465 &      2897.0308 &      2166.3382  \\
 7 &      1149.8200 &     -2743.1742 &     -8497.3906 &      5252.0887 &     -1220.4556 &      -673.3240  \\
 8 &      -268.0686 &       630.9120 &      1929.4613 &     -1182.3208 &       268.9864 &       102.7608  \\
 9 &        25.9818 &       -60.1922 &      -181.3145 &       109.9412 &       -23.9653 &        -5.3795  \\
\hline
\hline
$i$ & $\chi_{b1}(1P)$ & $\chi_{b1}(2P)$ & $\chi_{b1}(3P)$ &
      $\chi_{b1}(4P)$ & $\chi_{b1}(5P)$ & $\chi_{b1}(6P)$  \\
\hline
 1 &        33.0397 &       239.7994 &      -164.4051 &        34.1495 &        26.9515 &         8.4967  \\
 2 &      -332.6519 &     -2634.9469 &      1721.8106 &      -485.7810 &      -126.6739 &       149.8510  \\
 3 &      1385.1827 &     10912.2385 &     -7107.3970 &      2029.1900 &       430.9352 &      -885.8464  \\
 4 &     -2958.4499 &    -23096.8480 &     15028.0903 &     -4151.1836 &     -1094.3863 &      1678.5809  \\
 5 &      3605.9605 &     27851.8047 &    -18066.4365 &      4843.3153 &      1580.6111 &     -1375.8433  \\
 6 &     -2602.3368 &    -19853.0333 &     12821.3870 &     -3359.3446 &     -1243.8826 &       421.1791  \\
 7 &      1097.2744 &      8252.3736 &     -5300.9385 &      1361.1611 &       531.5536 &        55.6201  \\
 8 &      -249.2044 &     -1843.7114 &      1176.8983 &      -295.8920 &      -115.6049 &       -59.6685  \\
 9 &        23.5136 &       170.6412 &      -108.1134 &        26.4999 &         9.9105 &         9.4863  \\
\hline
\hline
$i$ & $h_b(1P)$ & $h_b(2P)$ & $h_b(3P)$ &
      $h_b(4P)$ & $h_b(5P)$ & $h_b(6P)$  \\
\hline
 1 &       -26.9406 &       245.4519 &      -158.2896 &        29.3812 &        24.3700 &        10.9592 \\
 2 &       265.1898 &     -2696.0191 &      1654.8622 &      -435.3349 &       -96.8333 &       123.8420 \\
 3 &     -1104.2180 &     11165.1311 &     -6832.1653 &      1822.9779 &       306.2952 &      -785.0079 \\
 4 &      2358.6741 &    -23627.7741 &     14450.3845 &     -3718.3930 &      -838.9663 &      1479.3562 \\
 5 &     -2874.8697 &     28482.8717 &    -17375.2937 &      4327.6261 &      1290.5769 &     -1149.1916 \\
 6 &      2074.5354 &    -20294.0610 &     12332.5098 &     -2998.2763 &     -1052.4158 &       266.7467 \\
 7 &      -874.6420 &      8431.1444 &     -5099.3191 &      1214.5985 &       458.5851 &       117.4293 \\
 8 &       198.6257 &     -1882.4288 &      1132.2297 &      -264.0975 &      -100.7958 &       -72.9291 \\
 9 &       -18.7411 &       174.0887 &      -104.0177 &        23.6613 &         8.6824 &        10.6504 \\
\hline
\end{tabular}
}
\end{center}
\end{table}

\newpage
\begin{figure}
\centerline{\epsfig{file=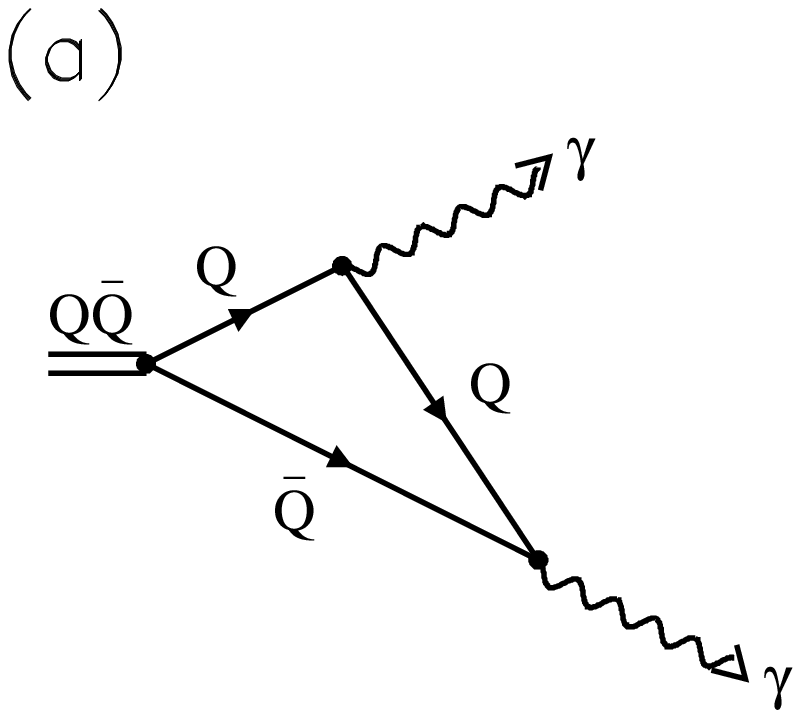,width=6cm}\hspace{0cm}
            \epsfig{file=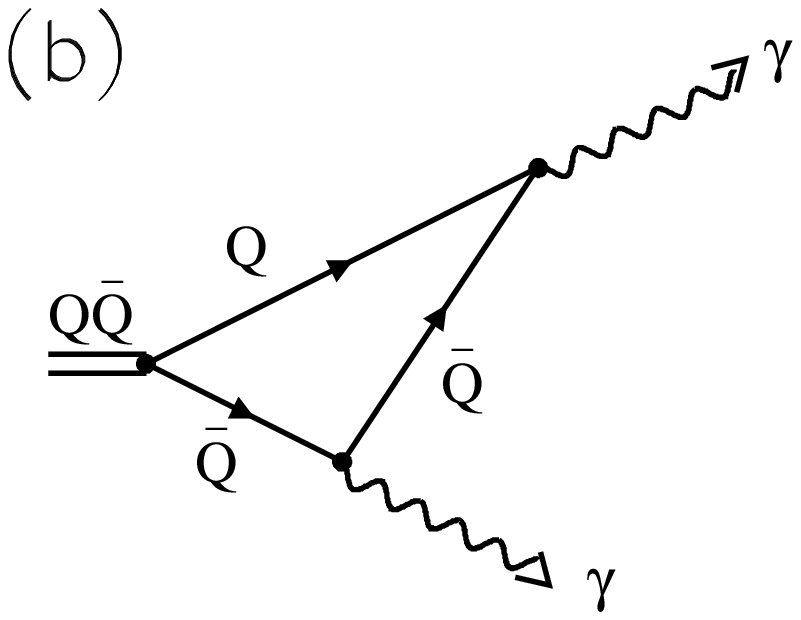,width=6cm}\hspace{0cm}
            \epsfig{file=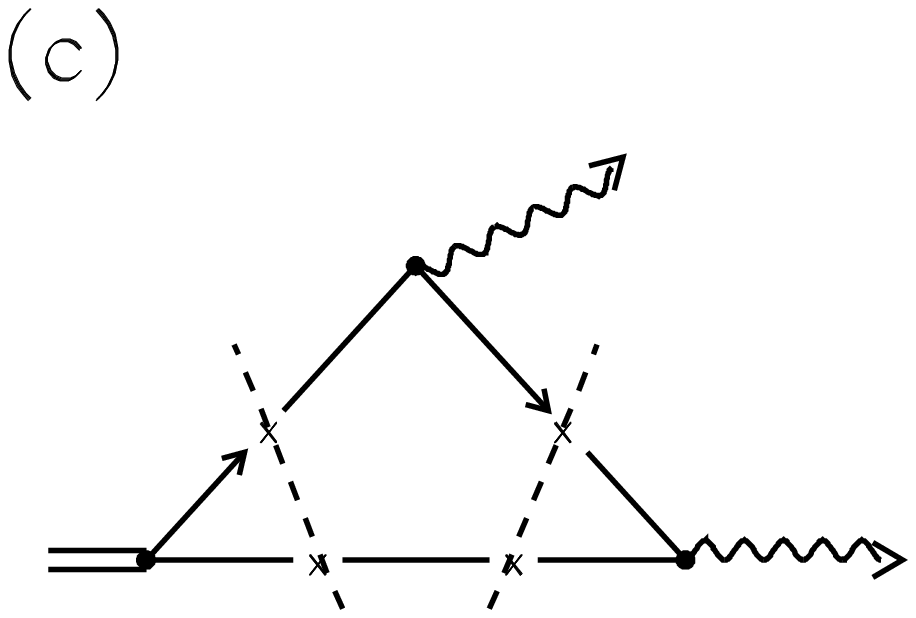,width=6cm}}
\centerline{\epsfig{file=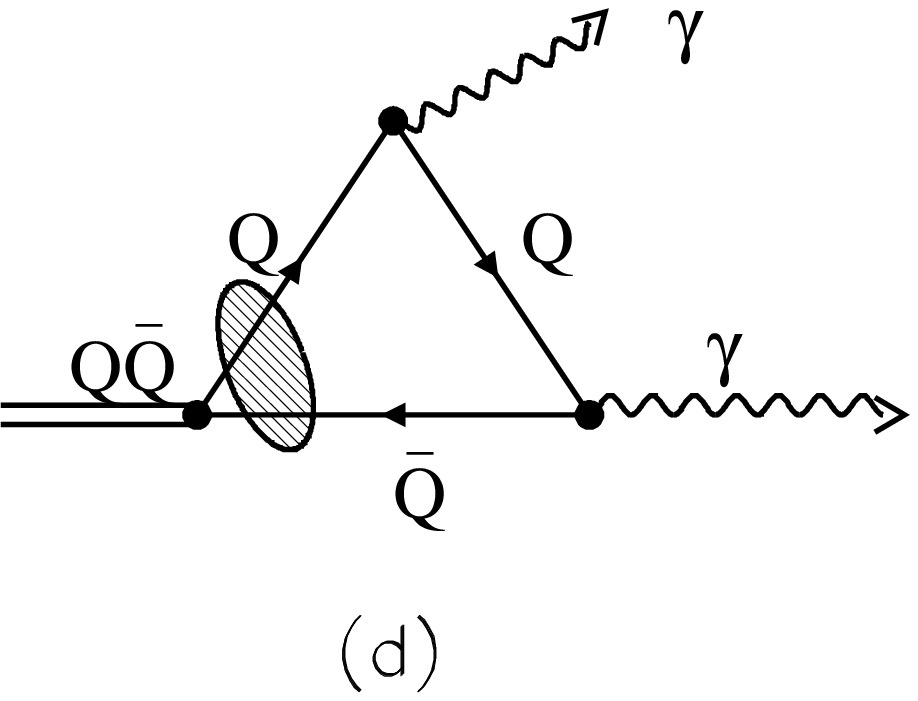,width=6cm}\hspace{1cm}
            \epsfig{file=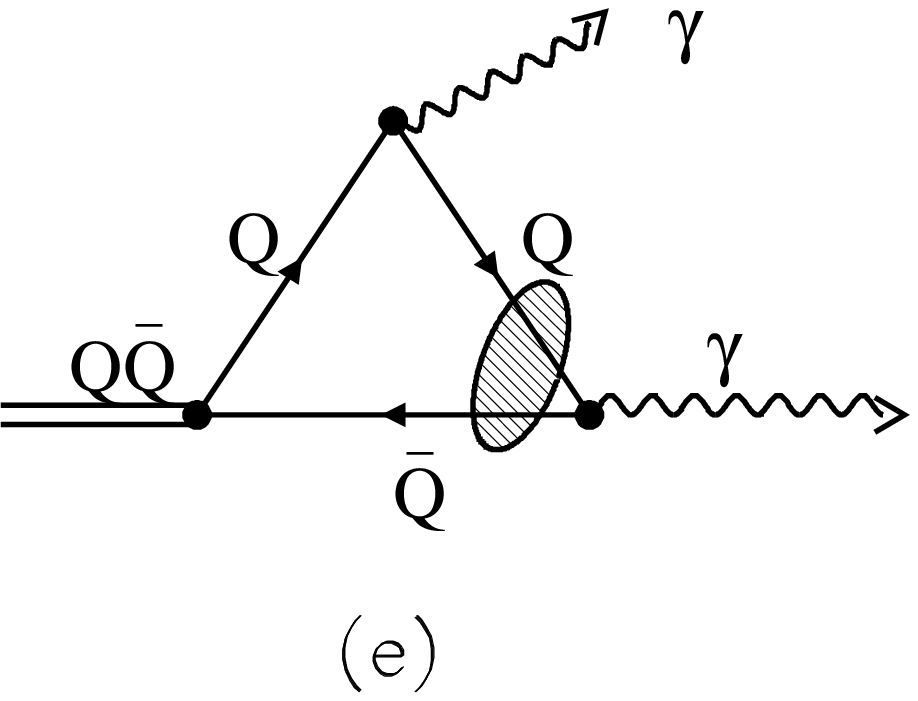,width=6cm}}
\centerline{\epsfig{file=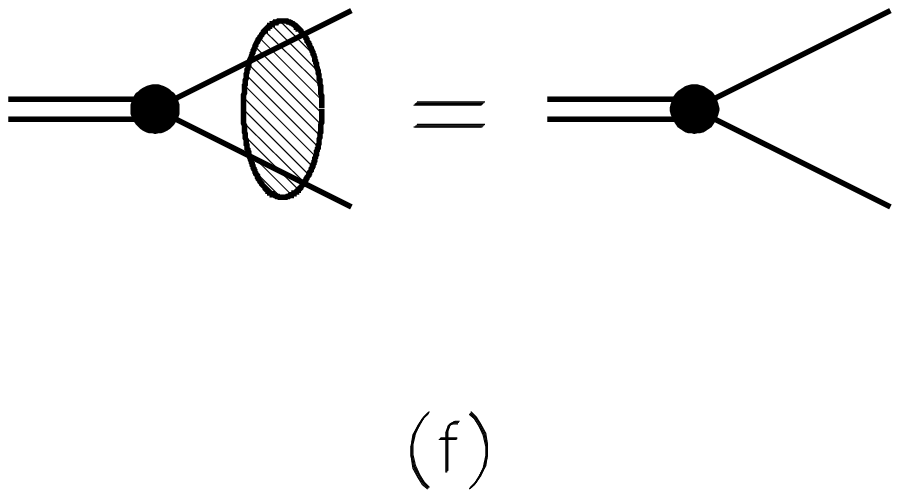,width=6cm}\hspace{1cm}
            \epsfig{file=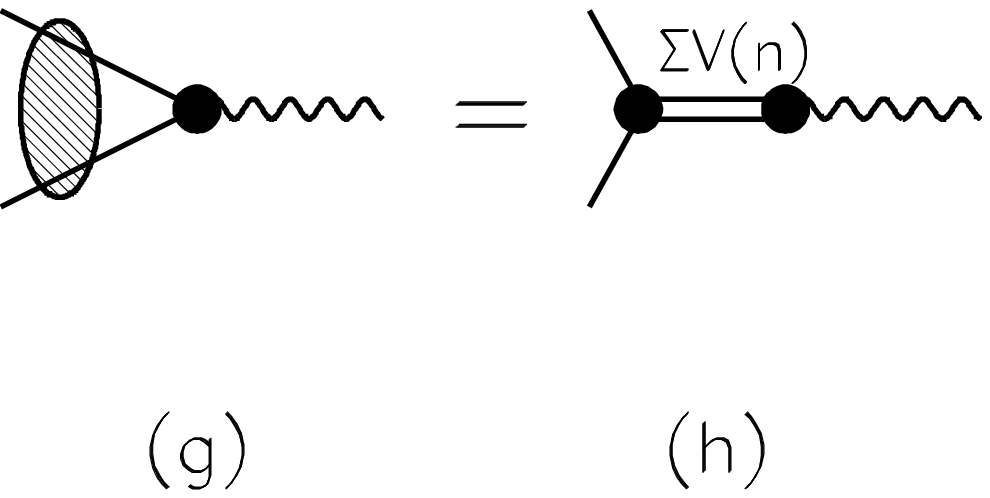,width=6cm}}
\vspace{0.5cm}
\caption{a,b) Diagrams for the two-photon decay of $Q\bar Q$ state and
c) cuttings in the spectral integral representation.
Initial (d) and final (e) state interactions of quarks in the
decay diagrams.
f) Graphical representation of the
spectral integral equation for the $Q\bar Q$ vertex. g,h)  Interaction
of
 quarks in the vertex $Q\bar Q\to\gamma $ and its
approximation by the sum of transitions
$Q\bar Q \to \sum\limits_{n}\,V(n) \to \gamma $ .}
\end{figure}

\begin{figure}
\centerline{\epsfig{file=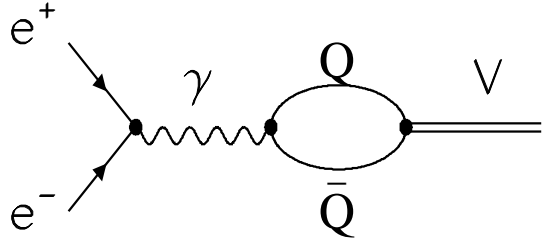,width=9cm}}
\vspace{0.5cm}
\caption{Quark transition diagram for the process $e^+e^-\to V$ .}
\end{figure}

\begin{figure}
\centerline{\epsfig{file=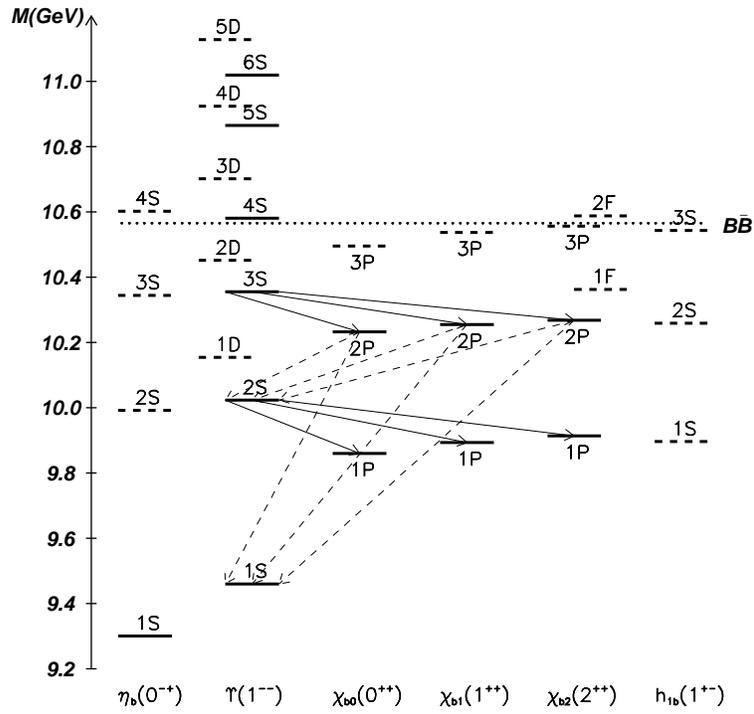,width=10cm}}
\vspace{1.0cm}
\caption{Radiative decays of the bottomonium systems, which
were taken into account in the fit (solid lines).
The dashed lines show radiative transitions with the known ratios for the
branchings
$Br[\chi_{bJ}(2P)\to\gamma\Upsilon(2S)]/
Br[\chi_{bJ}(2P)\to\gamma\Upsilon(1S)]$ ,
 these ratios are not included in the fit.}
\end{figure}

\begin{figure}
\centerline{\epsfig{file=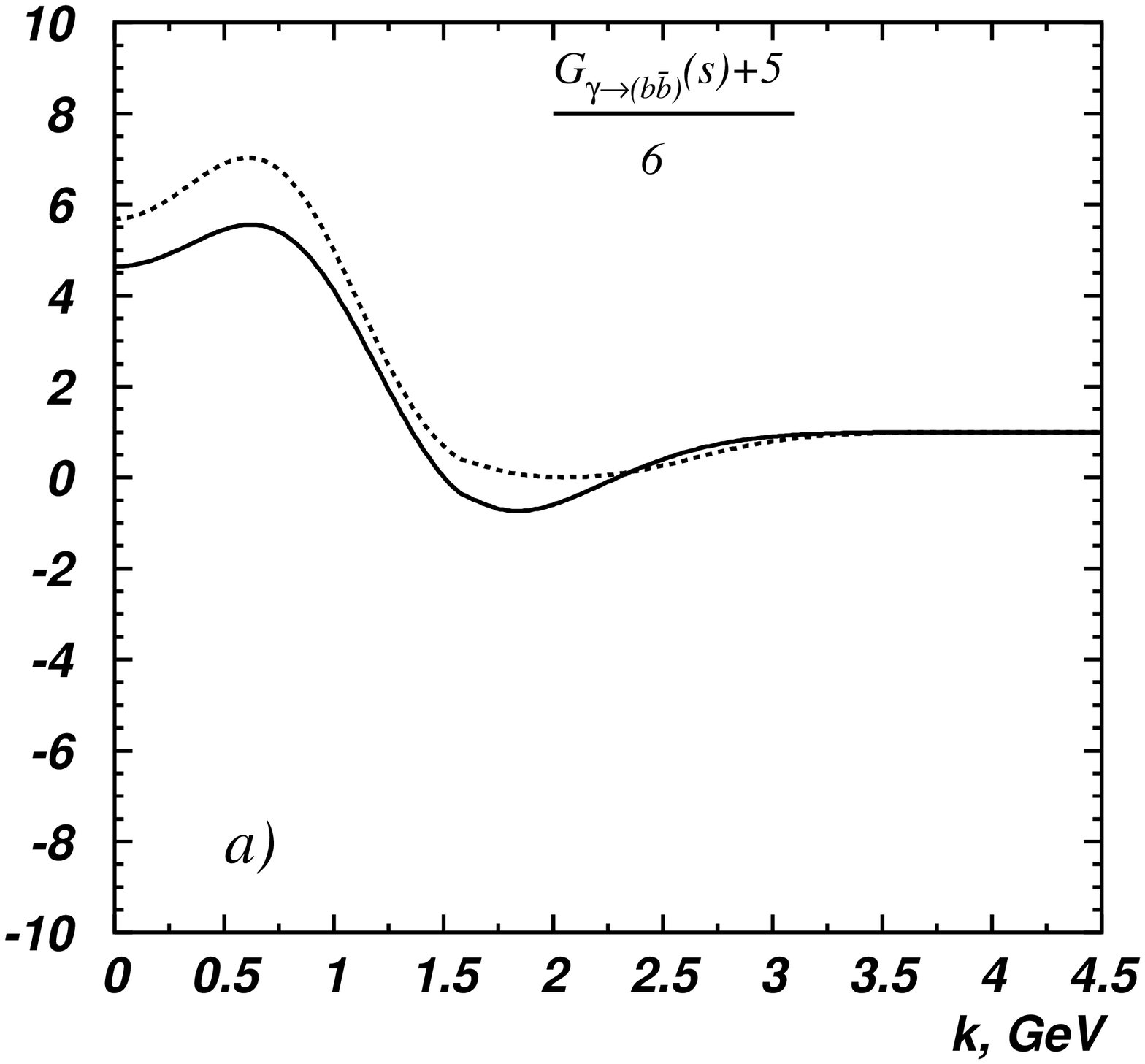,width=8cm}
            \epsfig{file=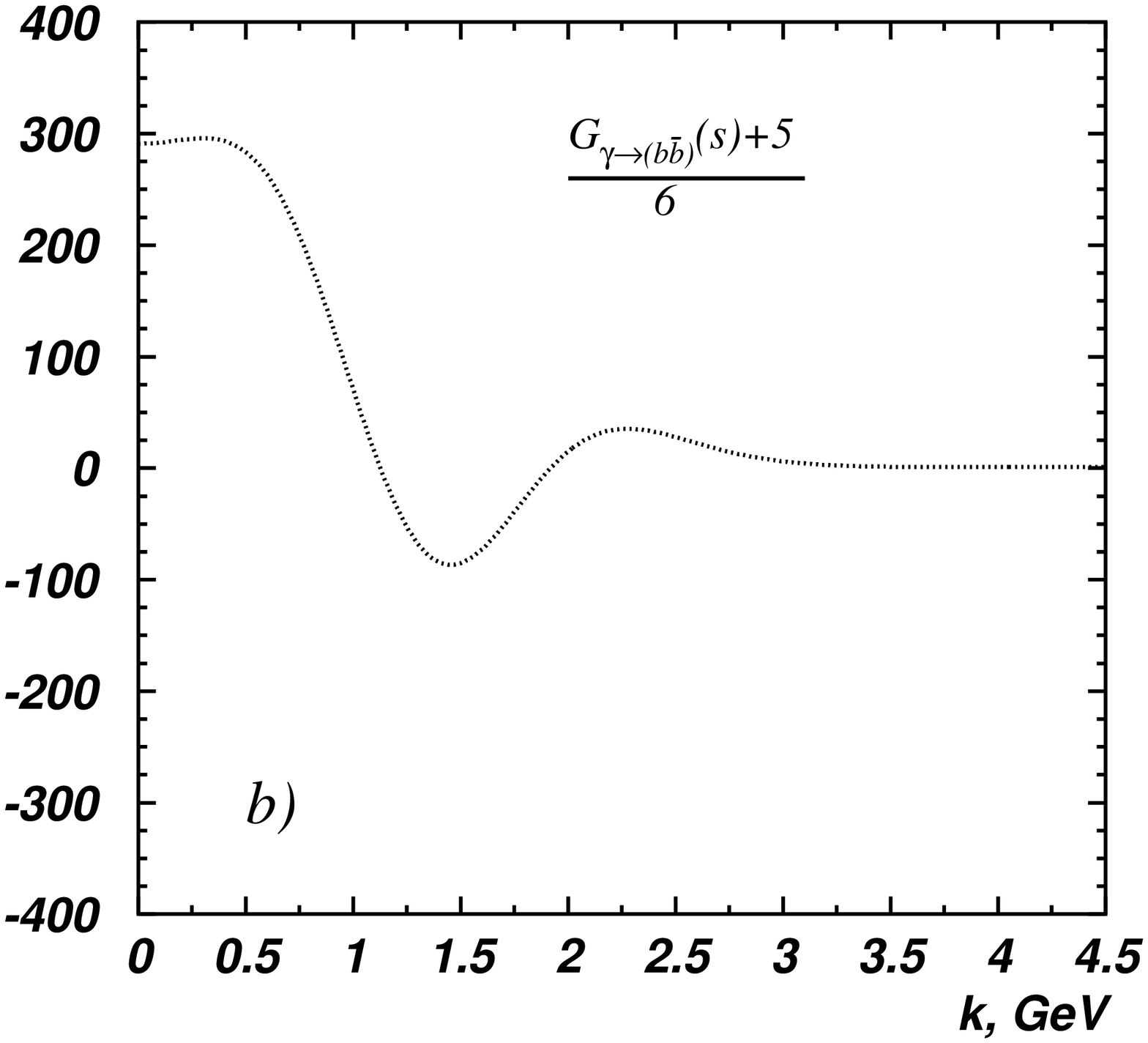,width=8cm}}
\centerline{\epsfig{file=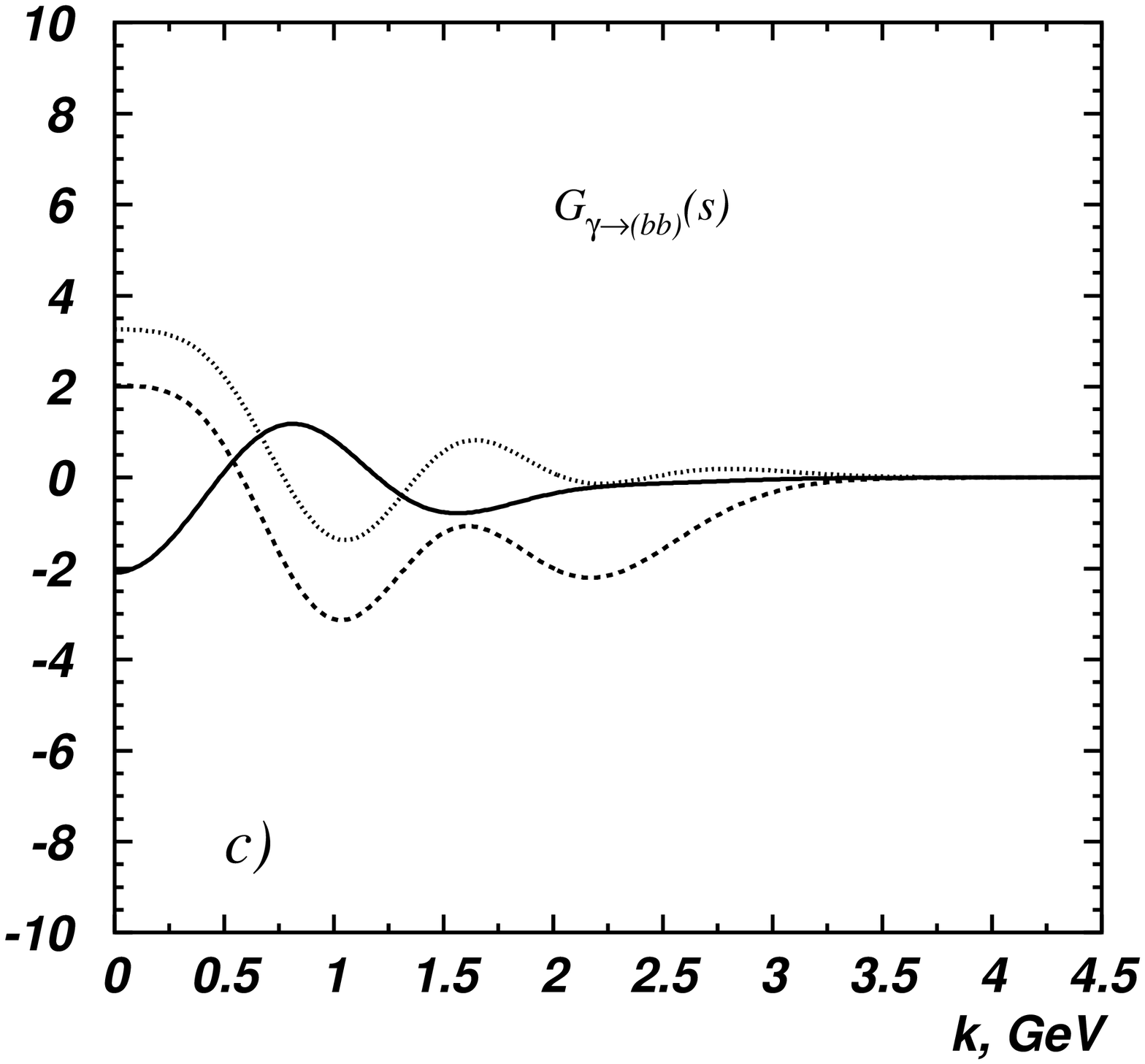,width=8cm}}
\caption{
a) Vertices $G^{(S)}_{\gamma\to b\bar b}$ (a) and
$G^{(D)}_{\gamma\to b\bar b}$ (b)
for the solutions $I(b\bar b)$ (solid line), $U(b\bar b)$ (dash line)
and $R(b\bar b)$ (dotted line).}
\end{figure}

\begin{figure}[h]
\centerline{\epsfig{file=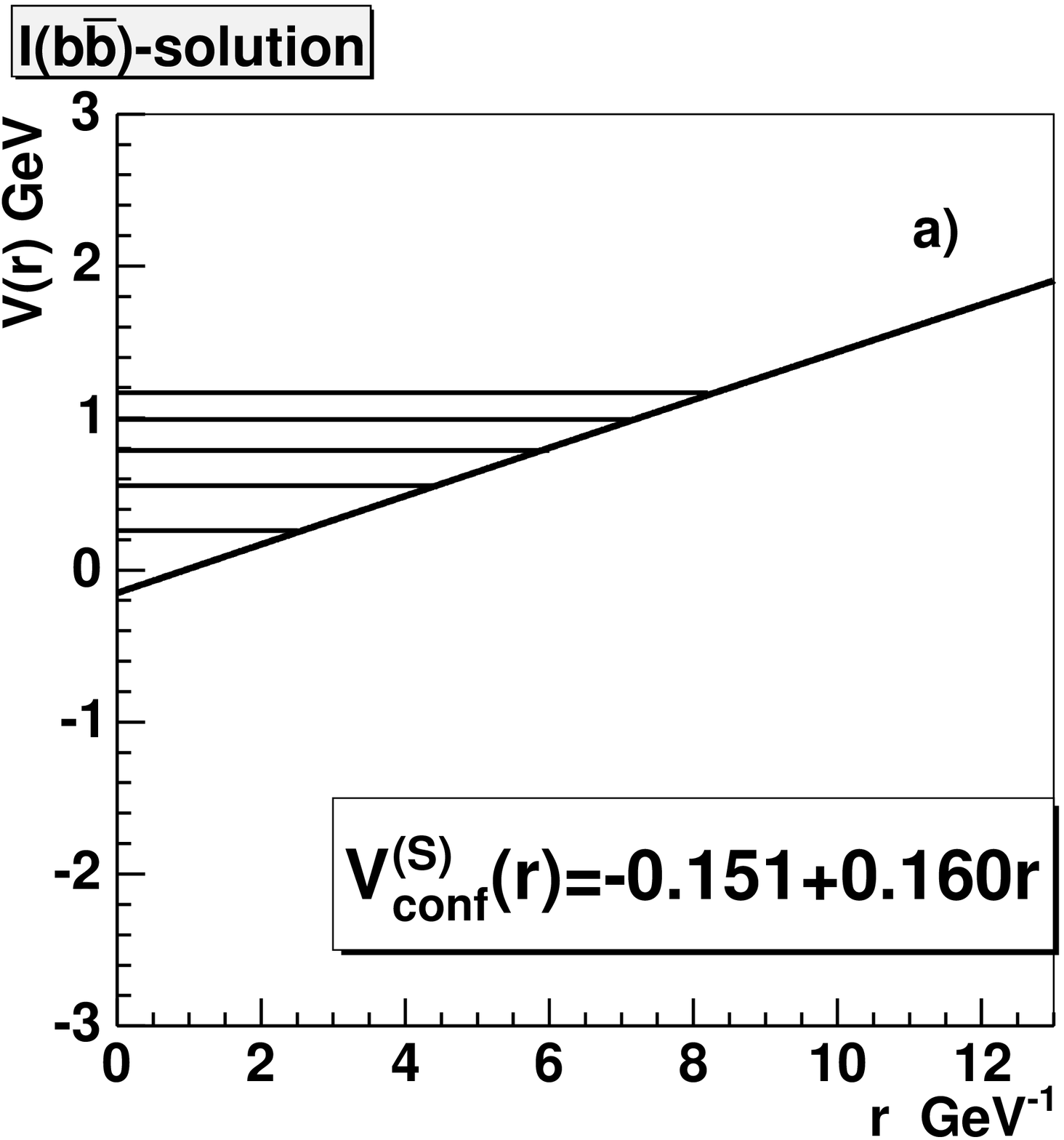 ,width=9cm}\hspace{0cm}
            \epsfig{file=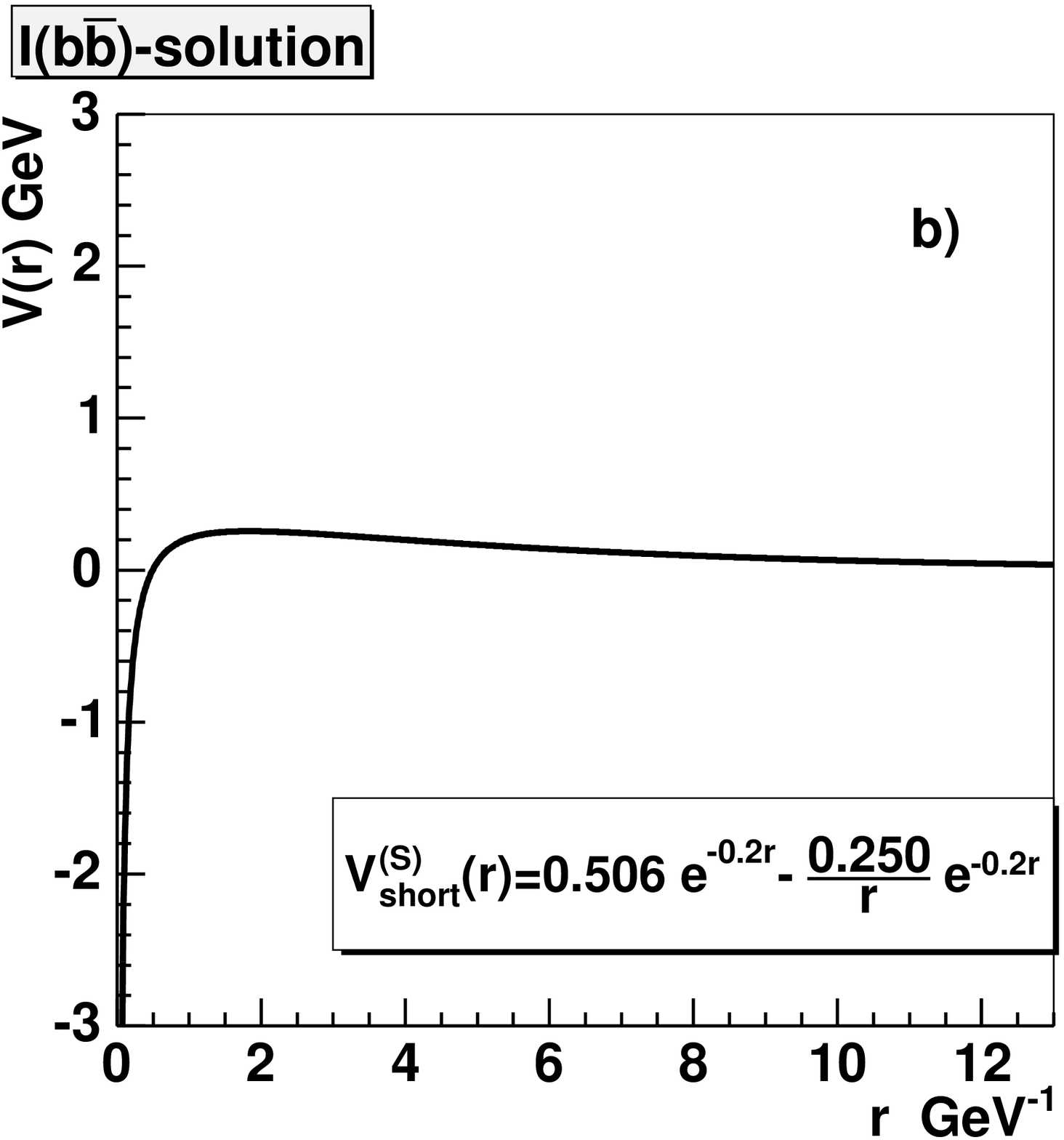,width=9cm}}
\centerline{\epsfig{file=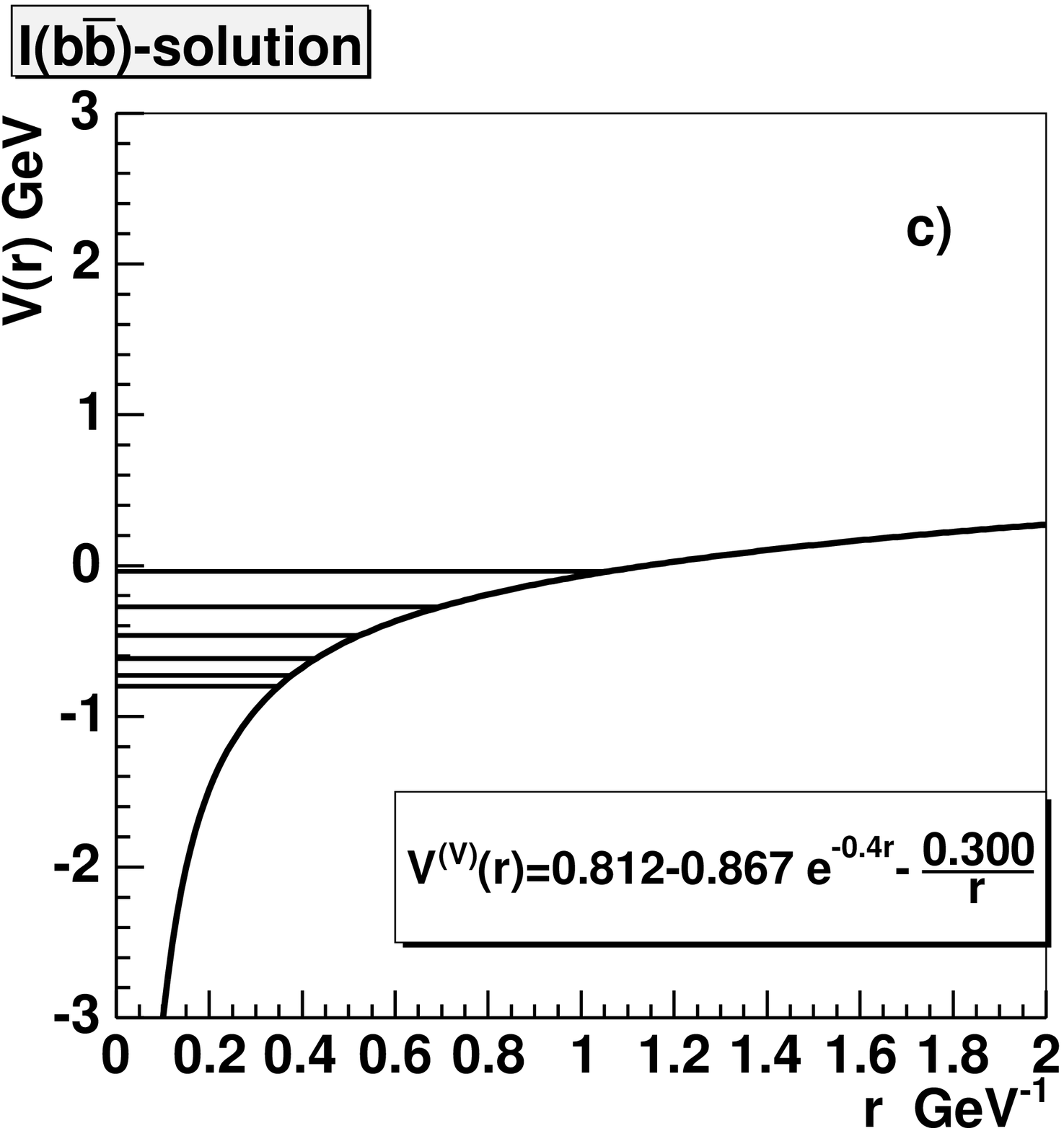 ,width=9cm}}
\caption{Potential for the solution $I(b\bar b)$: confinement potential
(a), short-range repulsive one (b) and attractive potential due to
vector exchange (c). To be illustrative, we show the $0^{-+}$ levels,
which would be created by the attractive potentials.} \end{figure}

\begin{figure}
\centerline{\epsfig{file=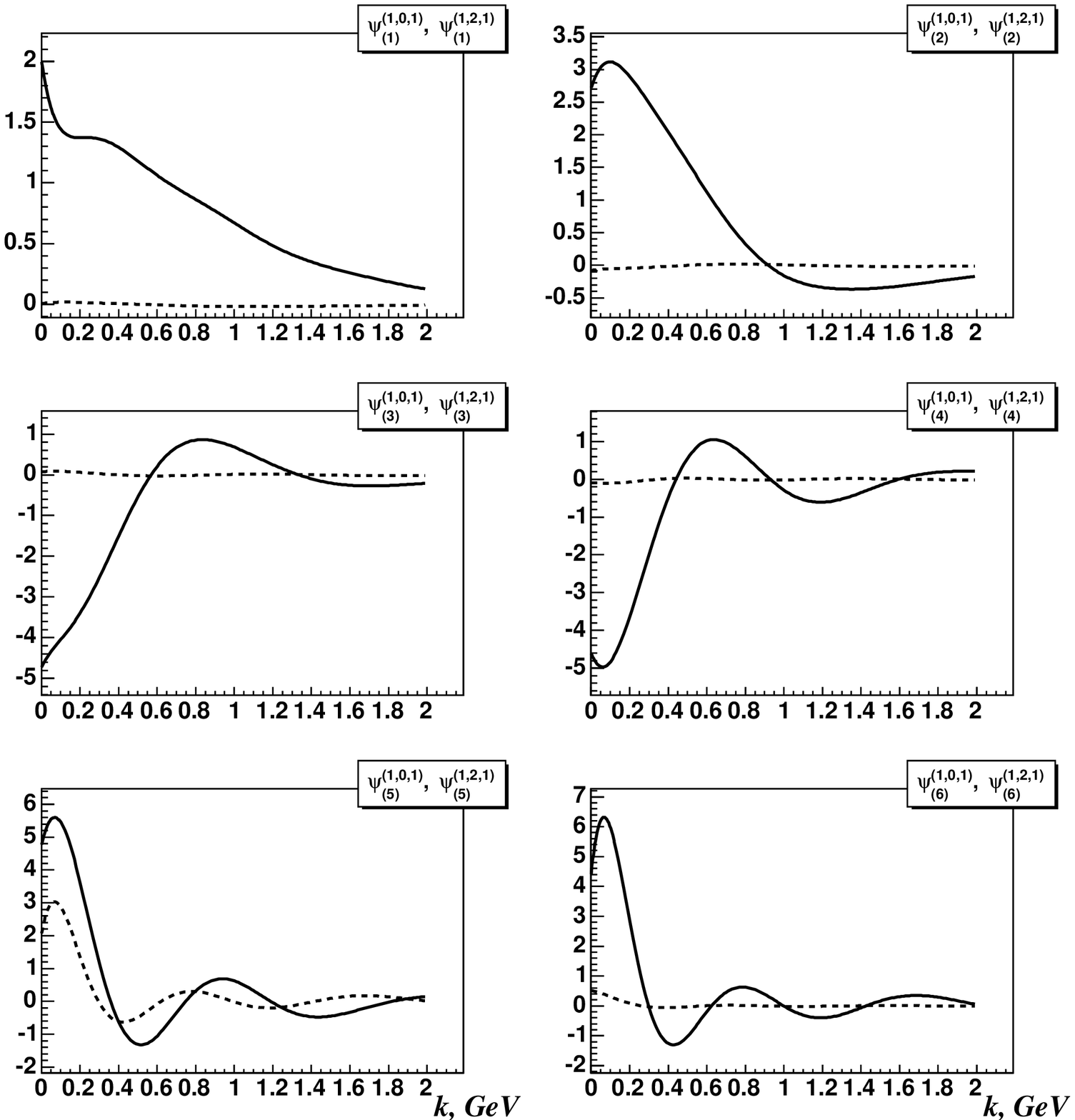,width=15cm}}
\caption{Wave functions for $\Upsilon(nS)$ in the solution $I(b\bar b)$.
Solid and dashed lines stand for $\psi^{(1,0,1)}$ and $\psi^{(1,2,1)}$.}
 \end{figure}

\begin{figure}
\centerline{\epsfig{file=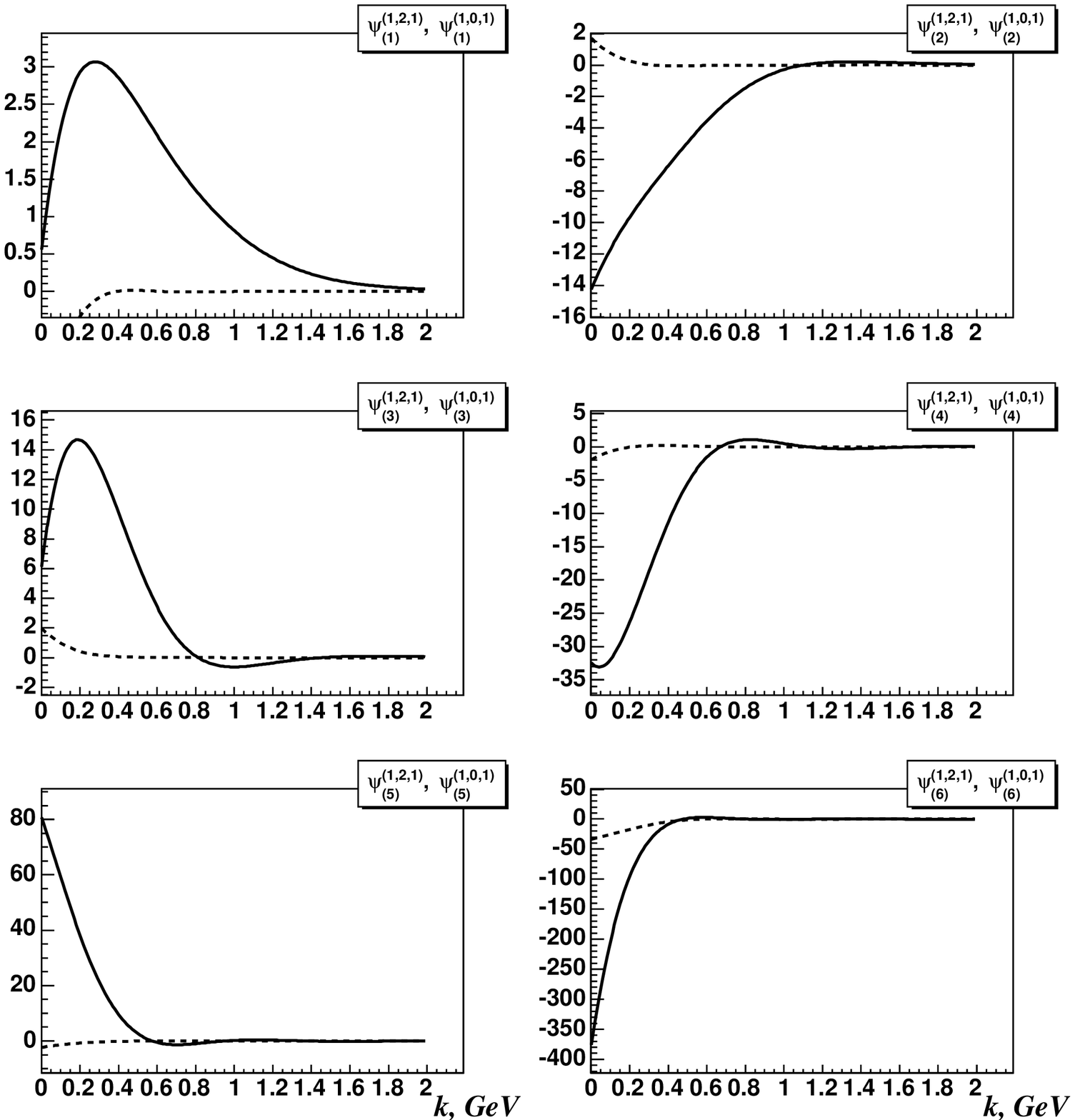,width=15cm}}
\caption{Wave functions for $\Upsilon(nD)$ in the solution $I(b\bar
b)$. Solid and dashed lines stand for $\psi^{(1,0,1)}$ and
$\psi^{(1,2,1)}$.}
 \end{figure}

\begin{figure}
\centerline{\epsfig{file=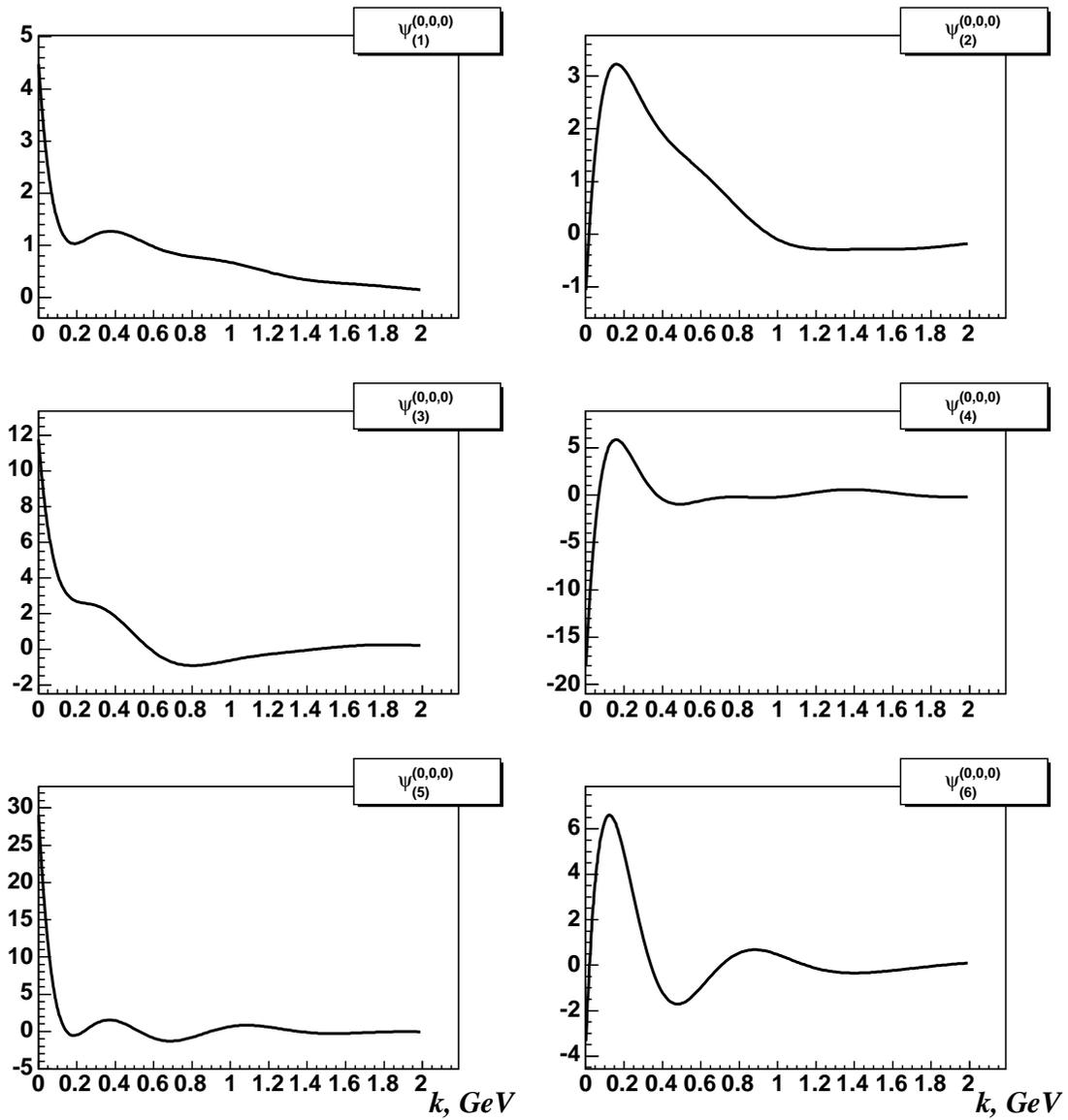,width=15cm}}
\caption{Wave functions for $\eta_{b0}$ in the solution $I(b\bar b)$.}
 \end{figure}

\begin{figure}
\centerline{\epsfig{file=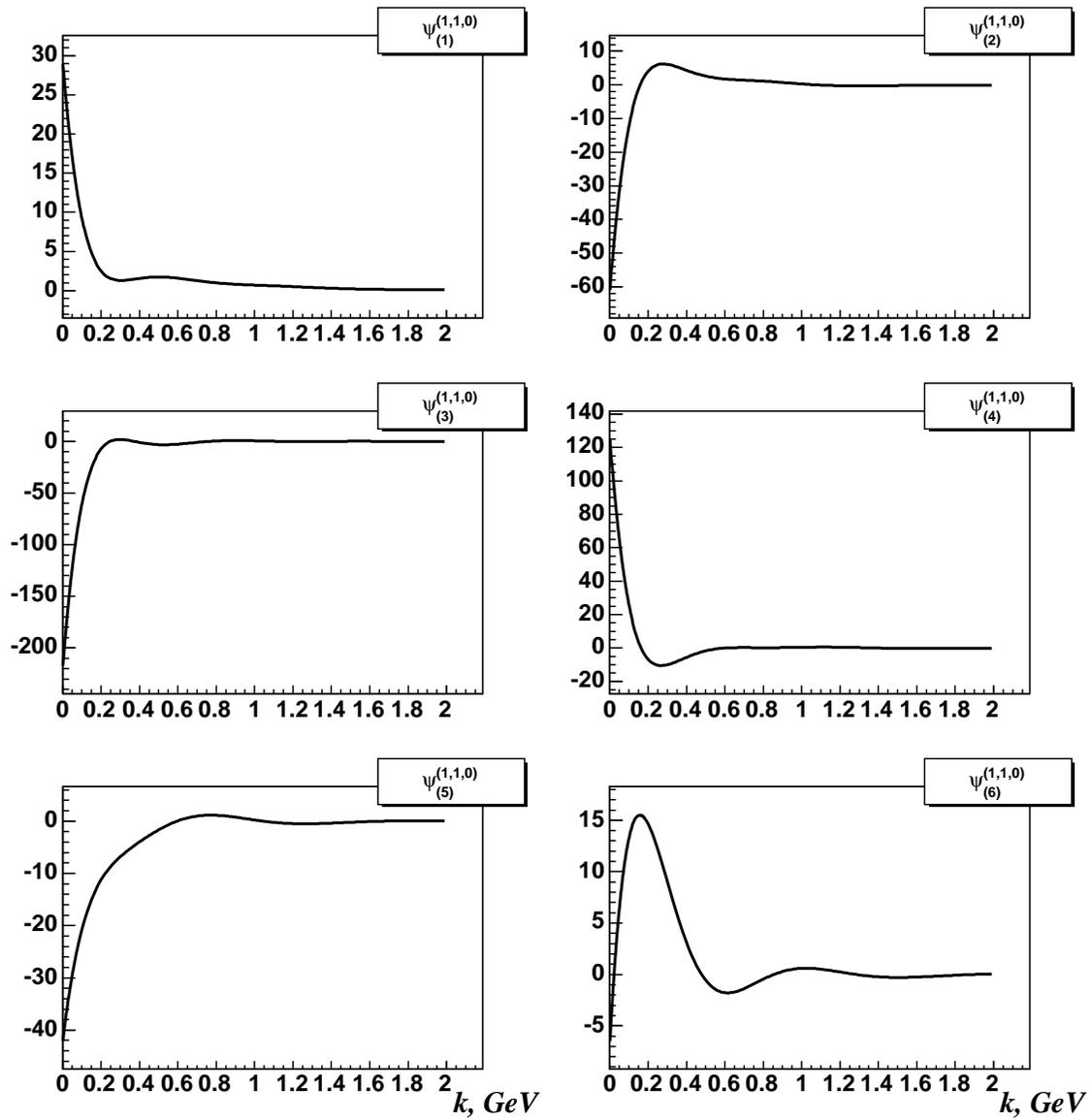,width=15cm}}
\caption{Wave functions for $\chi_{b0}$ in the solution $I(b\bar b)$.}
 \end{figure}

\begin{figure}
\centerline{\epsfig{file=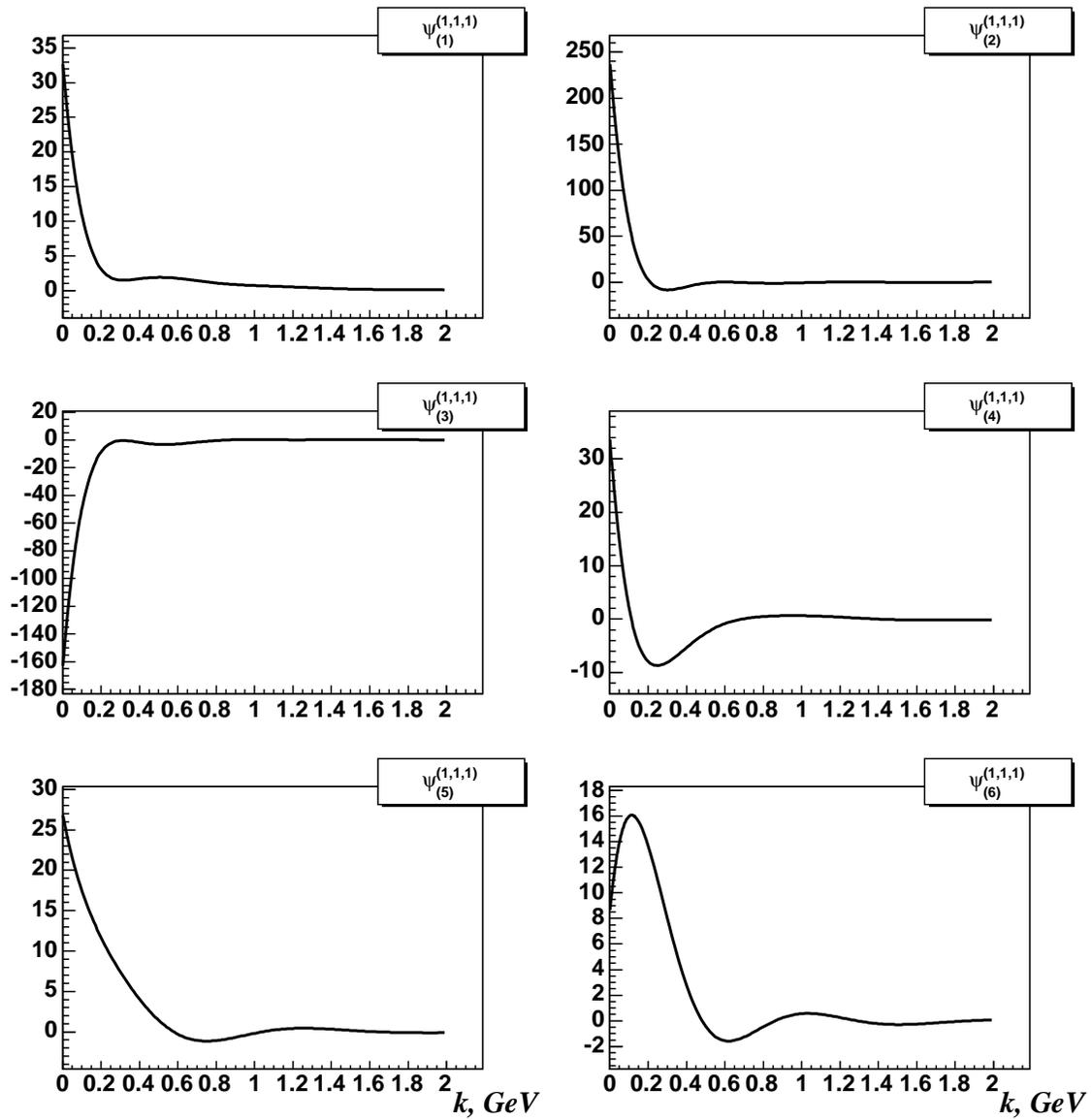,width=15cm}}
\caption{Wave functions for $\chi_{b1}$ for the solutions $I(b\bar
b)$.}
 \end{figure}

\begin{figure}
\centerline{\epsfig{file=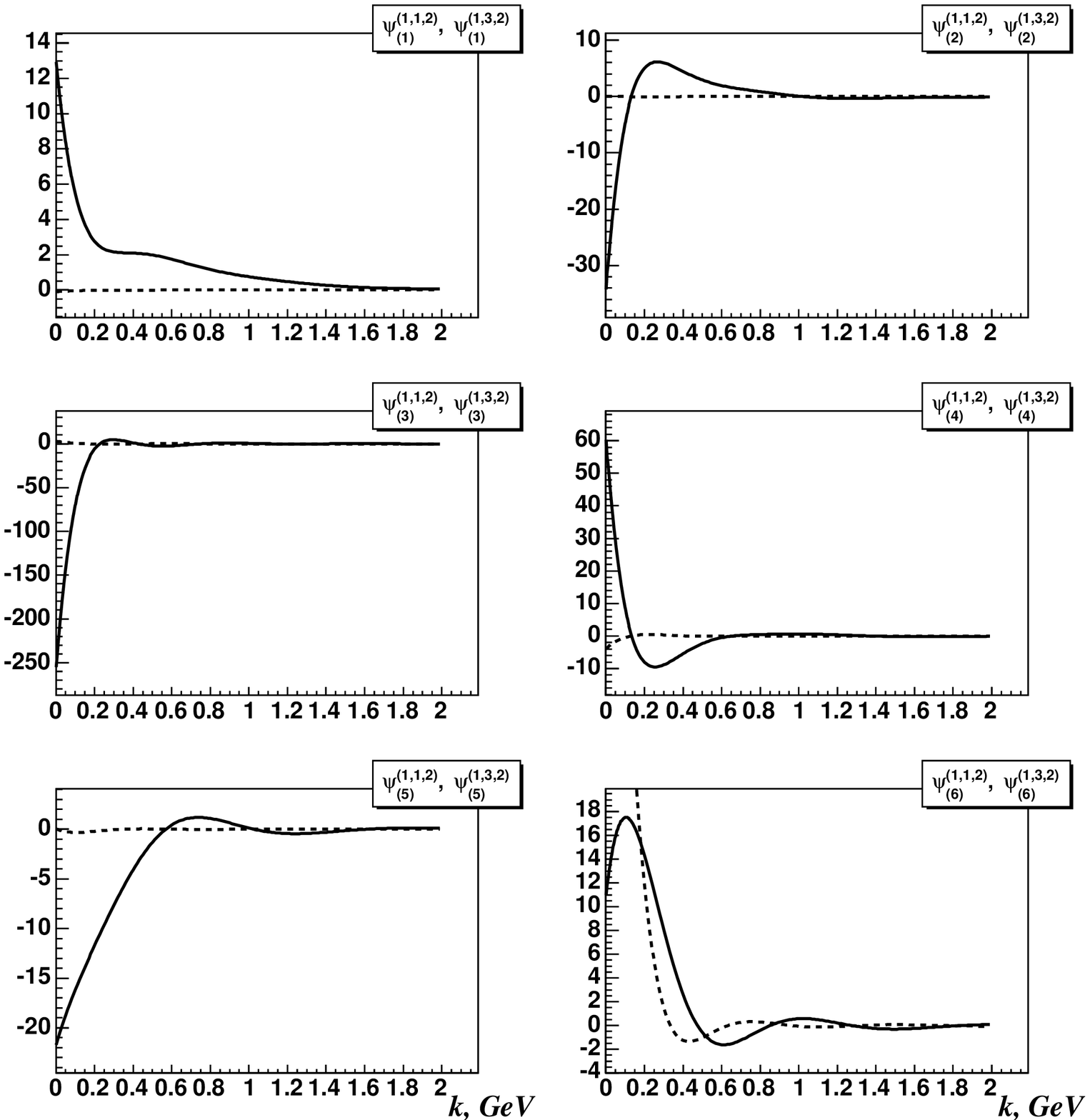,width=15cm}}
\caption{Wave functions for $\chi_{b2}$ in the solution $I(b\bar b)$.
Solid and dashed lines stand for $\psi^{(1,1,2)}$ and $\psi^{(1,3,2)}$.}
 \end{figure}

\begin{figure}
\centerline{\epsfig{file=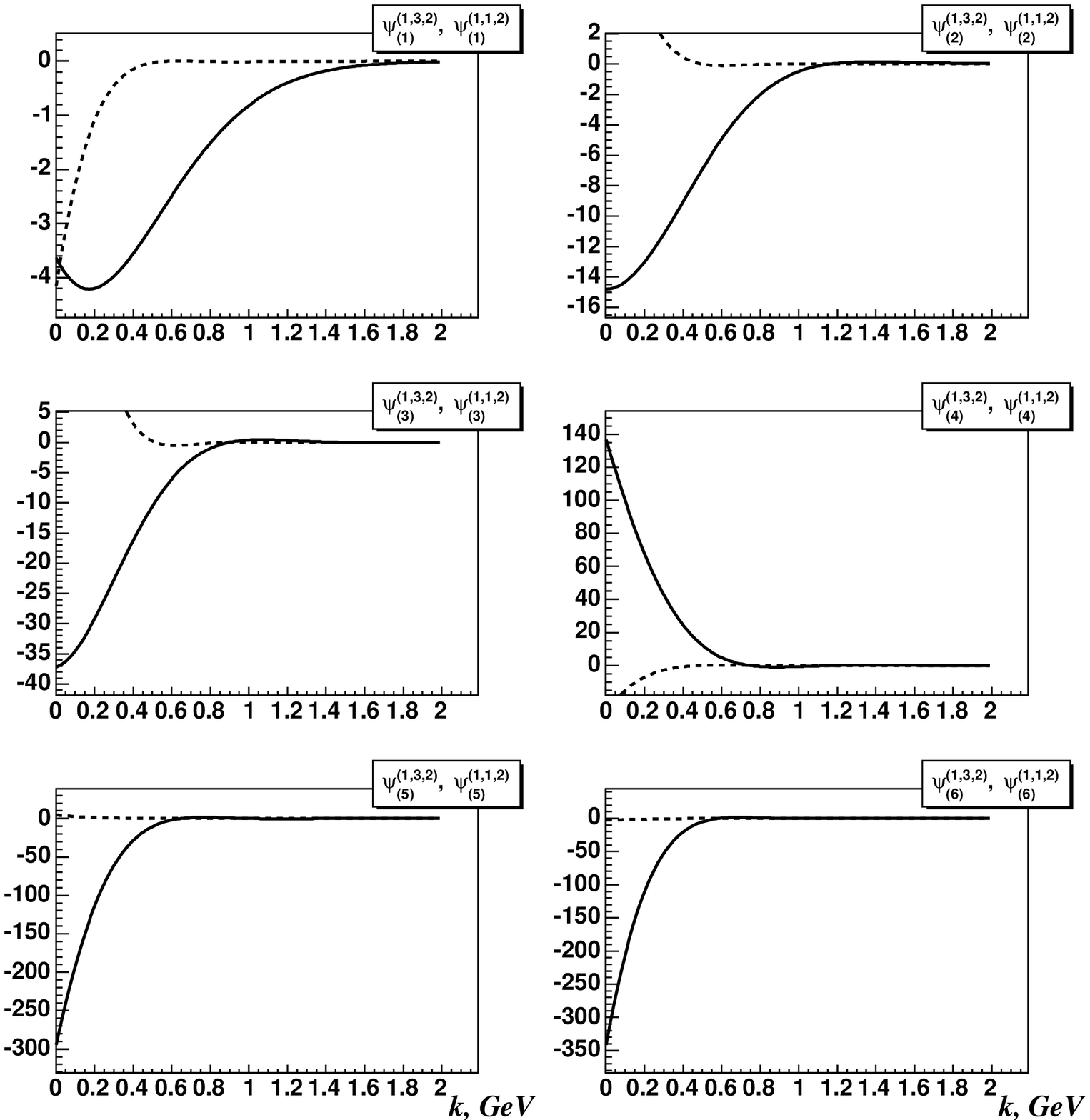,width=15cm}}
\caption{Wave functions for $\chi_{b2}$ in Solution $I(b\bar b)$.
Solid and dashed lines stand for $\psi^{(1,3,2)}$ and $\psi^{(1,1,2)}$.}
 \end{figure}

\begin{figure}
\centerline{\epsfig{file=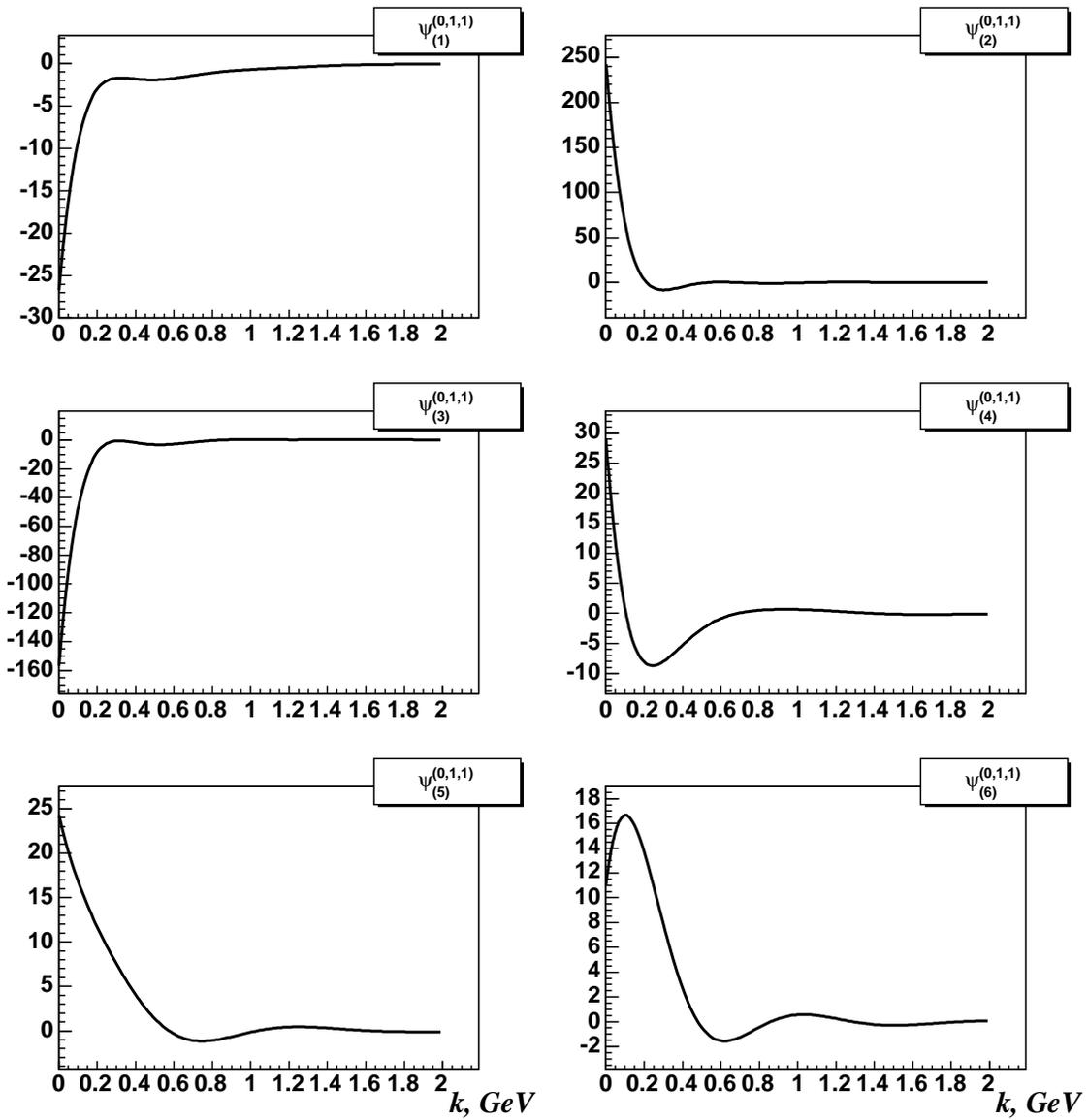,width=15cm}}
\caption{Wave functions for $b_{b1}$ in the solutions $I(b\bar b)$.}
 \end{figure}

\end{document}